\documentclass[11pt,a4paper]{article}
\usepackage{jheppub}
\usepackage{amsmath,amssymb,bm}
\usepackage{graphicx}
\usepackage{hepunits}
\usepackage{braket}
\usepackage{slashed}

\usepackage{subfigure}

\DeclareMathOperator{\Tr}{Tr}
\DeclareMathOperator{\Imag}{Im}

\allowdisplaybreaks

\newcommand{\ff}{f\!\!f}

\title{Resummation of soft and Coulomb corrections for $t\bar{t}h$ production at the LHC}

\author[a]{Wan-li Ju}
\emailAdd{wanli\_ju@pku.edu.cn}
\author[a,b,c]{and Li Lin Yang}
\emailAdd{yanglilin@pku.edu.cn}

\affiliation[a]{School of Physics and State Key Laboratory of Nuclear Physics and Technology, Peking University, Beijing 100871, China}
\affiliation[b]{Collaborative Innovation Center of Quantum Matter, Beijing, China}
\affiliation[c]{Center for High Energy Physics, Peking University, Beijing 100871, China}

\abstract{In this paper, a combined resummation of soft and Coulomb corrections is performed for the associated production of the Higgs boson with a top quark pair at the LHC. We illustrate the similarities and critical differences between this process and the $t\bar{t}$ production process. We show that up to the next-to-leading power, the total cross section for $t\bar{t}h$ production admits a similar factorization formula in the threshold limit as that for $t\bar{t}$ production. This fact, however, is not expected to hold at higher powers. Based on the factorization formula, we perform the resummation at the improved next-to-leading logarithmic accuracy, and match to the next-to-leading order result. This allows us to give NLL$'$+NLO predictions for the total cross sections at the LHC. We find that the resummation effects enhance the NLO cross sections by about 6\%, and significantly reduce the scale dependence of the theoretical predictions.}

\begin{document}
 
\maketitle

\section{Introduction}
\label{sec:introduction}
 
After the discovery of the Higgs boson at the Large Hadron Collider (LHC) in 2012 \cite{Aad:2012tfa, Chatrchyan:2012xdj}, a main task of particle physics is to investigate its properties.
One particularly important property is the Yukawa coupling between the top quark and the Higgs boson, which is crucial to understand the origin of the large top quark mass.
Precise knowledge of the top quark Yukawa coupling will also help us to constrain new physics effects in other couplings such as the Higgs boson self-coupling \cite{Goertz:2014qta, Azatov:2015oxa, Cao:2015oaa, Shen:2015pha}.
At the LHC, the top-Higgs Yukawa coupling can be probed by measuring the cross section for Higgs boson production in association with a top quark pair ($t\bar{t}h$ production). The observation of this process has been established very recently by the ATLAS \cite{Aaboud:2018urx} and CMS \cite{Sirunyan:2018hoz} collaborations. The measured cross sections are in agreement with theoretical predictions, although there are still large experimental uncertainties due to limited statistics. In the future, more precise measurements will be carried out at the upgraded LHC and the High Luminosity phase of the LHC (HL-LHC) \cite{Apollinari:2015bam}. For this reason, it is necessary to improve the theoretical understanding of this process.

At the leading order (LO), $t\bar{t}h$ production can be initialized by $q\bar{q}$ or $gg$ from the colliding protons, which has been studied many years ago \cite{Ng:1983jm, Kunszt:1984ri, Dicus:1988cx}. The next-to-leading order (NLO) quantum chromodynamics (QCD) corrections were calculated in \cite{Beenakker:2001rj, Reina:2001bc, Reina:2001sf, Beenakker:2002nc, Dawson:2002tg, Dawson:2003zu}, while the NLO electroweak contributions were calculated in \cite{Yu:2014cka, Frixione:2014qaa, Frixione:2015zaa}.
Given the complicated structure of this scattering process, it will be very challenging to perform an exact calculation of the next-to-next-to-leading order corrections.
Therefore, a lot of efforts have been devoted to approximations of the higher order QCD corrections in various kinematic limits \cite{Kulesza:2015vda, Broggio:2015lya, Broggio:2016lfj, Kulesza:2017ukk}.
The derivation of an approximation often involves a factorization formula, which can be used to resum a class of large corrections to all orders in perturbation theory.
For example, Refs.~\cite{Broggio:2015lya, Broggio:2016lfj, Kulesza:2017ukk} have investigated the limit $\hat{s} \to M^2_{t\bar{t}h}$, where $\sqrt{\hat{s}}$ is the partonic center-of-mass energy and $M_{t\bar{t}h}$ is the invariant mass of the $t\bar{t}h$ system. 
In this limit, large logarithmic corrections are present at each order in perturbation theory. These corrections are resummed to all orders in the strong coupling $\alpha_s$ up to the next-to-next-to-leading logarithmic (NNLL) accuracy \cite{Broggio:2016lfj, Kulesza:2017ukk}. In this paper, we consider a different kinematic limit $\hat{s} \to 2m_t + m_h$, where $m_t$ and $m_h$ are the masses of the top quark and the Higgs boson, respectively.

The limit under consideration is a bit different from the limit taken in \cite{Broggio:2016lfj, Kulesza:2017ukk}. There are power-like corrections arising from exchanges of Coulomb gluons, besides logarithmic corrections coming from soft gluon emissions. This limit has been studied in \cite{Kulesza:2015vda}, where the soft gluon contributions are resummed, but the Coulomb gluon contributions are only incorporated at fixed-order. In this paper, we will derive a factorization formula which can resum simultaneously both kinds of higher-order corrections. The framework presented in this paper closely resembles that in $t\bar{t}$ production \cite{Beneke:2009ye, Beneke:2010da, Beneke:2011mq}. However, it should be emphasized that $t\bar{t}h$ production is more complicated than $t\bar{t}$ production. In particular, in the threshold limit, the $t\bar{t}$ pair has a non-vanishing transverse momentum given by the recoil against the extra Higgs boson. This will lead to more involved interplay between the soft and Coulomb gluons, as will be clear later. The derivation of the factorization formula therefore requires new analyses other than those in \cite{Beneke:2009ye, Beneke:2010da}.
It is also much more difficult to calculate the hard function describing the contributions from hard gluons with typical momentum scale around $\sqrt{\hat{s}}$.

The paper is organized as follows. In Sec.~\ref{sec:lo}, we use the analytic form of the LO partonic cross sections to analyze their behavior in the threshold limit. In Sec.~\ref{sec:factorization}, we present the derivation of the factorization and resummation formulas using effective field theory methods. In Sec.~\ref{sec:numerics}, we show the numeric results based on our resummation formula. We conclude in Sec.~\ref{sec:conclusion}.

\section{Analyses of leading order results}
\label{sec:lo}

The total cross section for inclusive $t\bar{t}h$ production at hadron colliders can be expressed as \cite{Collins:1989gx}
\begin{equation}
\sigma(s,m_t,m_h) = \sum_{i,j} \int^1_{\tau_{\text{min}}} d\tau \, \hat{\sigma}_{ij}(\tau,m_t,m_h,\mu_f) \, \ff_{ij}(\tau,\mu_f) \, ,
\label{eq:sigma}
\end{equation}
where the sums are over all the partons within the colliding hadrons, i.e, $i(j) \in \lbrace q,\bar{q},g \rbrace$; $\sqrt{s}$ is the center-of-mass energy of the collider; $m_{t}$ and $m_h$ are the top quark mass and the Higgs mass, respectively; $\mu_f$ is the factorization scale; and
\begin{equation}
\tau_{\text{min}} = \frac{(2m_t+m_h)^2}{s} \, .
\end{equation}

The above factorization formula involves the partonic cross section $\hat{\sigma}_{ij}$ and the effective parton luminosity function $\ff_{ij}$. The definition of the latter is
\begin{equation}
\ff_{ij}(\tau,\mu_f) = \int^1_{\tau} \frac{d\xi}{\xi} \, f_{i/N_1}(\xi,\mu_f) \, f_{j/N_2}(\tau/\xi,\mu_f) \, ,
\end{equation}
where $f_{i/N}$ is the non-perturbative parton distribution function (PDF) of the parton $i$ in the hadron $N$. It is universal and can be extracted from experimental data. The partonic cross section $\hat{\sigma}_{ij}$ can be calculated in perturbative QCD. In this work, we are interested in its behavior near the threshold limit $\tau \gtrsim \tau_{\text{min}}$, or $\sqrt{\hat{s}} \gtrsim 2m_t + m_h$. Here $\sqrt{\hat{s}}$ is the partonic center-of-mass energy defined by $\hat{s} \equiv \tau s$. To analyze this region, we define $\beta=\sqrt{1-(2m_t+m_h)^2/\hat{s}}$. The parameter $\beta$ goes to zero in the threshold limit, and represents the typical momenta of final state particles. To see this, consider the energy conservation condition in the partonic center-of-mass frame
\begin{equation}
\sqrt{\hat{s}} = \sqrt{m_t^2+\vec{p}_t^2}+\sqrt{m_t^2+\vec{p}^2_{\bar{t}}}+\sqrt{m_h^2+\vec{p}^2_h}+E_X \, ,
\end{equation}
where $\vec{p}_{t(\bar{t})}$ and $\vec{p}_h$ are the 3-momenta of the (anti-)top quark and the Higgs boson, respectively, and $E_X$ is the total energy of other emitted particles in the final state.
In the limit $\beta \rightarrow 0$, the 3-momenta of the (anti-)top quark and the Higgs boson becomes much smaller than their rest mass. The right side of the above equation can then be expanded in the small momenta and we obtain the following relation:
\begin{equation}
\frac{\vec{p}_t^2}{2m_t}+\frac{\vec{p}_{\bar{t}}^2}{2m_t}+\frac{\vec{p}_h^2}{2m_h}+E_X \sim \sqrt{\hat{s}}-2m_t-m_h \sim \frac{\sqrt{\hat{s}}}{2}\beta^2 \, .
\end{equation}
We therefore have the power-counting
\begin{equation}
\sqrt{\hat{s}} \sim m_t \sim m_h \, , \quad |\vec{p}_{t,\bar{t},h}| \sim \sqrt{\hat{s}} \beta \, , \quad  E_X \sim \sqrt{\hat{s}} \beta^2 \, .
\label{eq:ScalingRules1}
\end{equation}
This will be important for establishing the effective field theory description later in the next section. For the moment, we are going to investigate the behavior of the partonic cross sections in the limit $\beta \to 0$.
 
The Born-level partonic cross section can be written as
\begin{equation}
\hat{\sigma}^{(0)}_{ij \to t\bar{t}h} = \frac{1}{2\hat{s}} \int \frac{d^3\vec{p}_t}{(2\pi)^3 2E_t} \, \frac{d^3\vec{p}_{\bar{t}}}{(2\pi)^3 2E_{\bar{t}}} \, \frac{d^3\vec{p}_h}{(2\pi)^3 2E_h} \, (2\pi)^4 \delta^4(p_i+p_j-p_t-p_{\bar{t}}-p_h) \, \overline{\big|\mathcal{M}_{ij \to t\bar{t}h}\big|^2} \, .
\end{equation}
To the first order in $\beta$, one can approximate the $\delta$-functions in the above formula as
\begin{equation}
\delta\bigg(\sqrt{\hat{s}} - 2m_t - m_h - \frac{\vec{p}_t^2}{2m_t} - \frac{\vec{p}_{\bar{t}}^2}{2m_t} - \frac{\vec{p}_h^2}{2m_h} \bigg) \, \delta^3(\vec{p_t} + \vec{p}_{\bar{t}} + \vec{p}_h) \, ,
\end{equation}
and the squared-amplitudes can be expanded according to the counting in eq.~(\ref{eq:ScalingRules1}). The integrals over the 3-momenta can then be carried out, and we arrive at approximate expressions of the partonic cross sections in the following form
\begin{equation}
\hat{\sigma}^{(0)\mathrm{App}}_{q\bar{q}\rightarrow t\bar{t}h} = \frac{2 \beta^4 m_t^5 \alpha_s^2}{9 v^2 \, m_h^{3/2} \, (m_h+2m_t)^{7/2}} \, , \quad
\hat{\sigma}^{(0)\mathrm{App}}_{gg\rightarrow t\bar{t}h} = \frac{\beta^4 m_t \alpha_s^2 \, ( 2m_h^4 - 7m_h^2 m_t^2 + 14m_t^4 )}{192 v^2 \, m_h^{3/2} \, (m_h+2 m_t)^{7/2}} \, ,
\label{eq:LOappTTH}
\end{equation}
where $v$ is the vacuum expectation value of the Higgs field and $\alpha_s$ is  the strong coupling constant.

A crucial feature of eq.~(\ref{eq:LOappTTH}) is that the leading order partonic cross sections are proportional to $\beta^4$ in the threshold limit. This should be contrasted to the case of a similar process, $t\bar{t}$ production, where the threshold behavior is given by \cite{Czakon:2008cx},
\begin{equation}
\sigma^{(0)\text{App}}_{q\bar{q}\rightarrow t\bar{t}}= \frac{\pi   \alpha_s^2 \beta }{9  m_t^2} \, , \quad
\sigma^{(0)\text{App}}_{gg\rightarrow t\bar{t}} = \frac{7 \pi  \alpha_s^2 \beta }{192 m_t^2} \, .
\label{eq:LOappTT}
\end{equation}
That is, the Born partonic cross sections are linear in $\beta$ in the threshold limit. The different behaviors imply that the threshold region is less important in the case of $t\bar{t}h$ production than for $t\bar{t}$ production. In order to study these different behaviors more precisely, we will numerically compare the approximate results to the exact ones in the following.
For our numerical computations, we set $m_t=\unit{173.5}{\GeV}$, $m_h=\unit{125.09}{\GeV}$ and $v=\unit{246.22}{\GeV}$ \cite{Tanabashi:2018oca}. The strong coupling constant is evolved from the initial condition $\alpha_s(m_Z) = 0.1181$ to the renormalization scale $\mu_r = m_t + m_h/2$, where $m_Z$ is the mass of the $Z$ boson.

\begin{figure}[t!]
\centering
\subfigure[Results for $\hat{\sigma}^{(0)}_{q\bar{q}\rightarrow t\bar{t}h}$ as a function of $\beta$.]{
\begin{minipage}[t]{0.5\linewidth}
\centering
\includegraphics[scale=0.82]{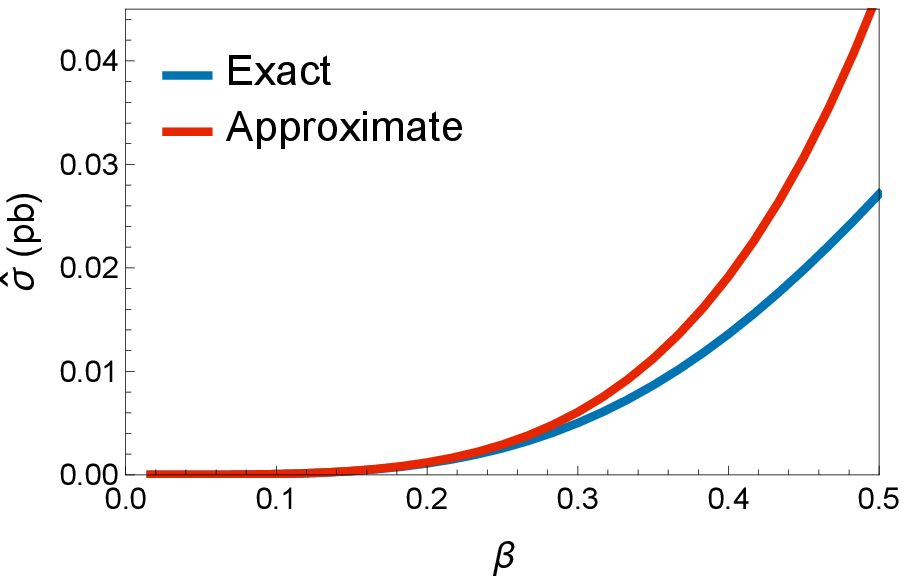}
\end{minipage}%
}%
\subfigure[Results for $\hat{\sigma}^{(0)}_{gg\rightarrow t\bar{t}h}$ as a function of $\beta$.]{
\begin{minipage}[t]{0.5\linewidth}
\centering
\includegraphics[scale=0.82]{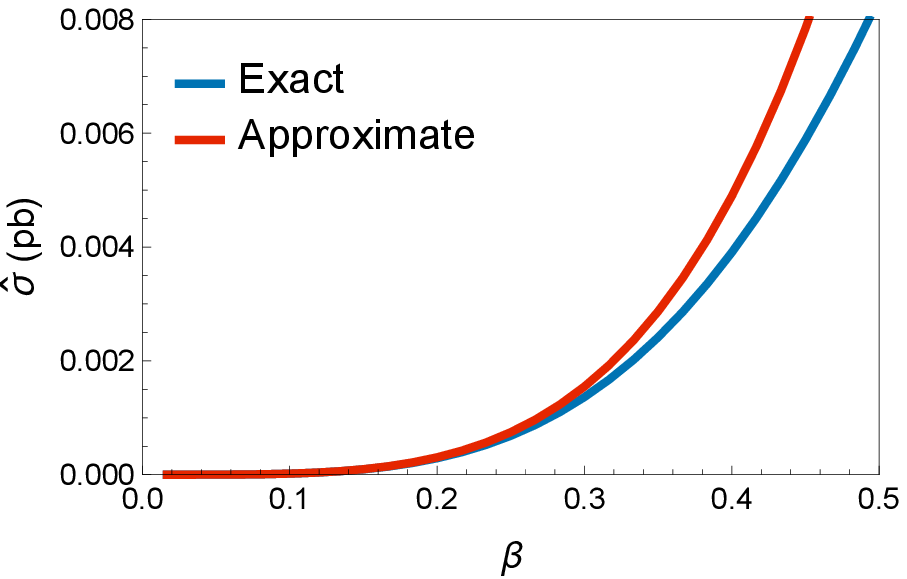}
\end{minipage}%
}%
\caption{Comparison between the exact and approximate Born partonic cross sections for $t\bar{t}h$ production.}
\label{fig:Born_QCD_app_individual_FO}
\end{figure}
 
In Fig.~\ref{fig:Born_QCD_app_individual_FO}, we numerically compare the approximate Born partonic cross sections for $t\bar{t}h$ production with the exact ones, i.e, $\hat{\sigma}^{(0)\text{Exact}}_{q\bar{q}(gg)\rightarrow t\bar{t}h}$.
The approximate results are obtained from Eq.~\eqref{eq:LOappTTH}.
To calculate $\hat{\sigma}^{ (0)\text{Exact}}_{q\bar{q}(gg)\rightarrow t\bar{t}h}$,  we first employ the programs FeynArts \cite{Hahn:2000kx} and FeynCalc \cite{Shtabovenko:2016sxi, Mertig:1990an} to generate the transition amplitudes and then use the Cuba library \cite{Hahn:2004fe, Hahn:2014fua} to perform the phase space integration. Our numerical results are checked against the automatic program MadGraph5\_aMC@NLO (MG5) \cite{Alwall:2014hca}.  
As shown in Fig.~\ref{fig:Born_QCD_app_individual_FO}, the approximate and exact partonic cross sections approach each other as $\beta$ becomes small. They are both highly suppressed in the threshold limit, as can be expected.
When $\beta$ grows larger, the approximate results start to overestimate the exact ones, indicating that there are important negative power corrections to the approximate formula.

\begin{figure}[t!]
\centering
\subfigure[Results for $\hat{\sigma}^{(0)}_{q\bar{q}\rightarrow t\bar{t}}$ as a function of $\beta$.]{
\begin{minipage}[t]{0.5\linewidth}
\centering
\includegraphics[scale=0.81]{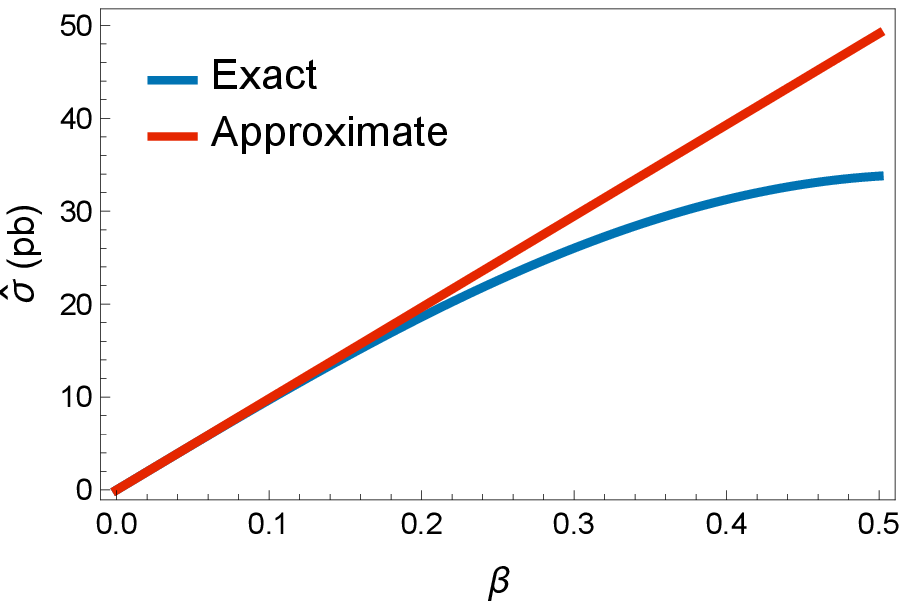}
\end{minipage}%
}%
\subfigure[Results for $\hat{\sigma}^{(0)}_{gg\rightarrow t\bar{t}}$ as a function of $\beta$.]{
\begin{minipage}[t]{0.5\linewidth}
\centering
\includegraphics[scale=0.81]{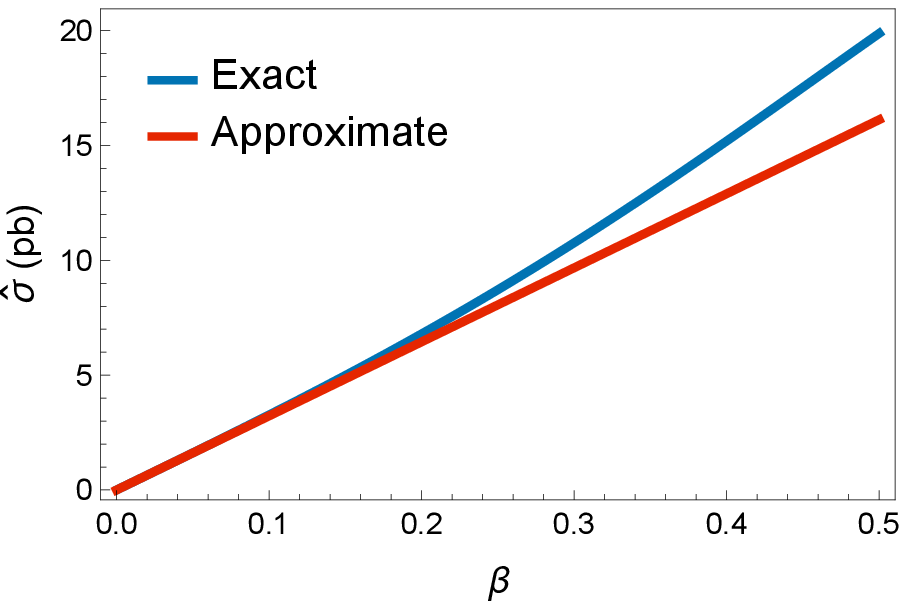}
\end{minipage}
}%
\caption{Comparison between the exact and approximate Born partonic cross sections for $t\bar{t}$ production.}
\label{fig:Born_QCD_app_ttbar}
\end{figure}

In Fig.~\ref{fig:Born_QCD_app_ttbar}, we show a similar comparison for the $t\bar{t}$ production process. The exact results are obtained from \cite{Moch:2008qy}, while the approximate results are computed using Eq.~\eqref{eq:LOappTT}. The first impression is that the small-$\beta$ region is less suppressed compared to the $t\bar{t}h$ case, which is clear from the $\beta^1$ vs. $\beta^4$ behaviors.
This means that the reliability of small-$\beta$ resummation can be quite different in these two processes, a fact not often mentioned in the literatures.
Beside this, here we also observe that the approximate result in the $q\bar{q}$ channel overestimates the exact one as $\beta$ grows. Interestingly, for the $gg \to t\bar{t}$ subprocess, the approximate result underestimates the exact one for most values of $\beta$, contrary to the $gg \to t\bar{t}h$ case shown in Fig.~\ref{fig:Born_QCD_app_individual_FO}.
It is therefore possible that, for the sum of the two channels, the approximate result stays closer to the exact one than in the case of individual channels. This should be investigated at the hadron level as we are going to do below. 

\begin{figure}[t!]
\centering
\subfigure[Results for $\mathrm{d}\sigma^{(0)}_{q\bar{q}(gg)\rightarrow t\bar{t}h}/\mathrm{d}\beta$.]{
\begin{minipage}[t]{0.5\linewidth}
\centering
\includegraphics[scale=0.82]{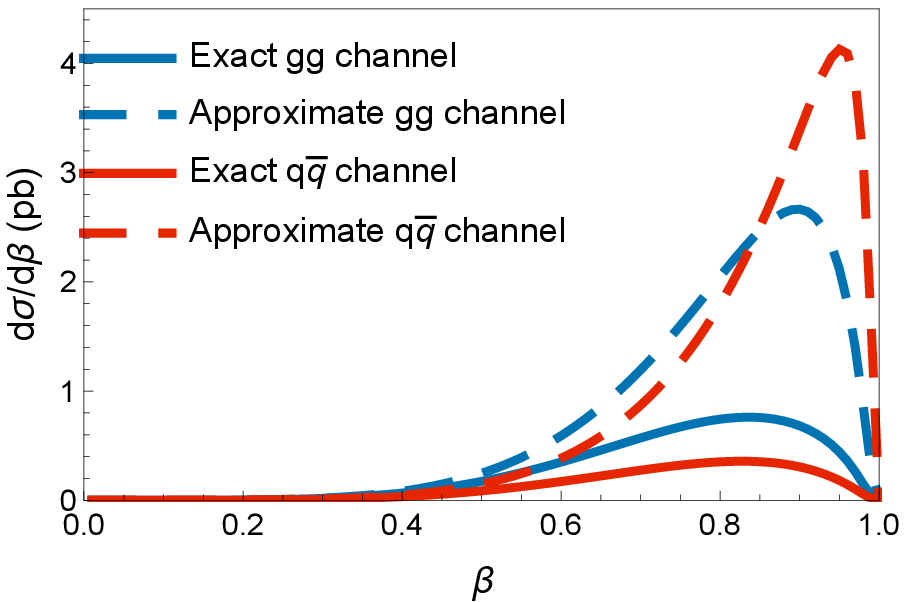}
\end{minipage}%
}%
\subfigure[Ratios between the approximate and exact results.]{
\begin{minipage}[t]{0.5\linewidth}
\centering
\includegraphics[scale=0.79]{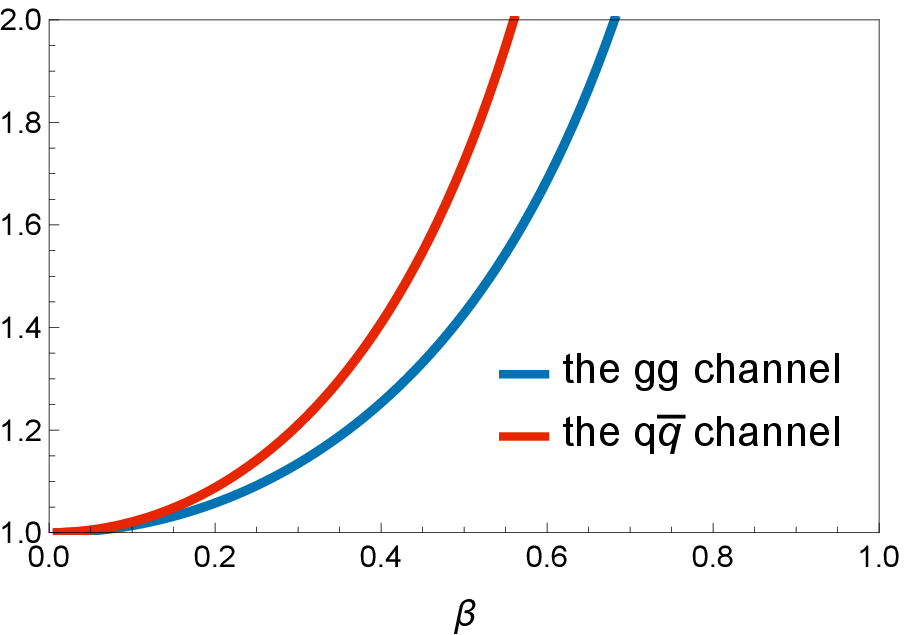}
\end{minipage}%
}%
\caption{The results for $\mathrm{d}\sigma^{(0)}_{q\bar{q}(gg)\rightarrow t\bar{t}h}/\mathrm{d}\beta$ at the 13 TeV LHC. The MMHT2014LO PDFs \cite{Harland-Lang:2014zoa} are employed here with the corresponding $\alpha_s(m_Z)$.}
\label{fig:dsigmadbeta_13TeV_pptth}
\end{figure}  

\begin{figure}[t!]
\centering
\subfigure[Results for $\mathrm{d}\sigma^{(0)}_{q\bar{q}(gg)\rightarrow t\bar{t}}/\mathrm{d}\beta$.]{
\begin{minipage}[t]{0.5\linewidth}
\centering
\includegraphics[scale=0.84]{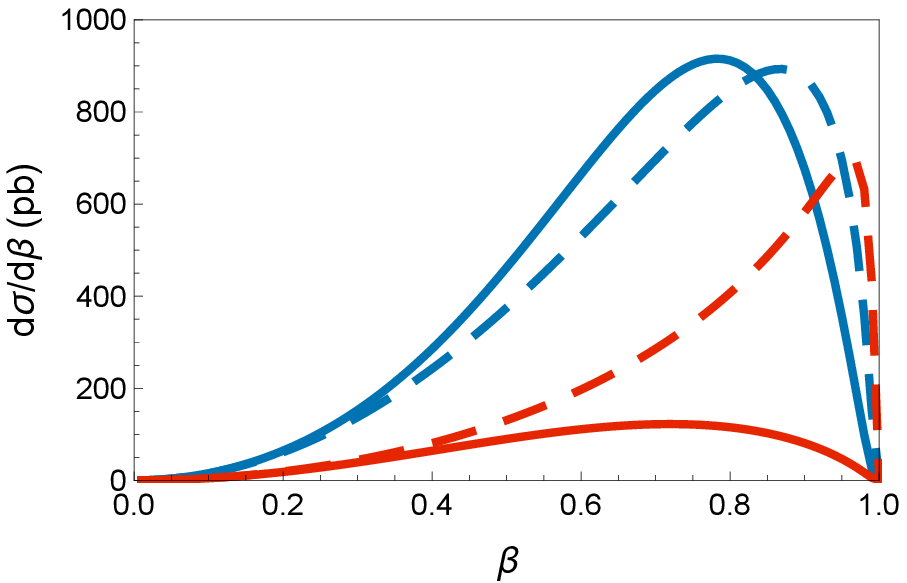}
\end{minipage}%
}%
\subfigure[Ratios between the approximate and exact results.]{
\begin{minipage}[t]{0.5\linewidth}
\centering
\includegraphics[scale=0.76]{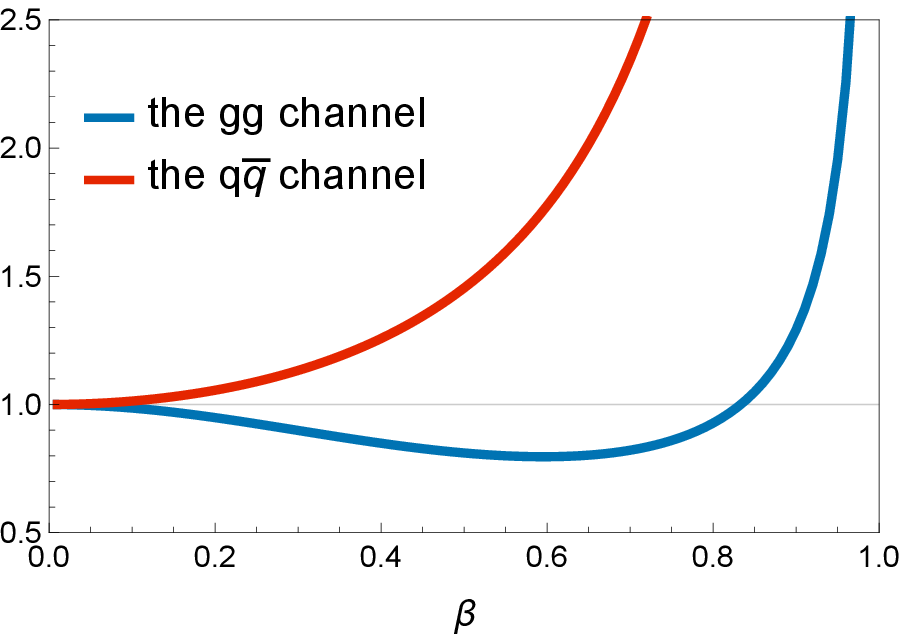}
\end{minipage}%
}%
\caption{The results for $\mathrm{d}\sigma^{(0)}_{q\bar{q}(gg)\rightarrow t\bar{t}}/\mathrm{d}\beta$ at the 13 TeV LHC. The notations are the same as Fig.~\ref{fig:dsigmadbeta_13TeV_pptth}.}
\label{fig:dsigmadbeta_13TeV_ppttbar}
\end{figure}  

The partonic cross sections need to be convoluted with the parton distribution functions (PDFs) to arrive at the hadronic cross sections. It is therefore interesting to see how the approximate and exact results compare at the hadron level. We show in Fig.~\ref{fig:dsigmadbeta_13TeV_pptth} the hadronic differential cross sections $d\sigma^{(0)}/d\beta$ for $t\bar{t}h$ production. From the left plot, we see again that the approximate results in both the $q\bar{q}$ and the $gg$ channels overshoot a lot over the exact ones for large $\beta$. It is also clear that the small-$\beta$ region is highly suppressed due to the $\beta^4$ behavior.
In the right plot, we show the ratio between the approximate results and the exact ones as a function of $\beta$. One can see that in the small-$\beta$ region, the approximate results are in good agreements with the exact ones. As $\beta$ goes above $\sim 0.3$, the approximation quickly fails. On the other hand, we show in Fig.~\ref{fig:dsigmadbeta_13TeV_ppttbar} the results for $t\bar{t}$ production. 
Again, we find that the small-$\beta$ region is more important in this case compared to $t\bar{t}h$ production due to the $\beta^1$ behavior. Besides, one can also see the accidental cancellation between the power corrections in the $q\bar{q}$ and $gg$ channel mentioned in the last paragraph. The above two facts lead to the observation that the small-$\beta$ expansion provides a reasonable approximation to the hadronic total cross section for $t\bar{t}$ production \cite{Bonciani:1998vc, Moch:2008qy, Beneke:2009ye, Beneke:2010da, Beneke:2011mq, Piclum:2018ndt}. This is clearly not the case for $t\bar{t}h$ production.

While the above analyses show that the small-$\beta$ region for $t\bar{t}h$ production is not as important as that for $t\bar{t}$ production, it is still interesting to study the small-$\beta$ behavior of the cross section at higher orders in QCD. First of all, theoretically, $t\bar{t}h$ production is similar but slightly different from $t\bar{t}$ production. The $t\bar{t}$ pair is recoiled against the Higgs boson and therefore has a non-vanishing transverse momentum already at the lowest order in QCD. The interplay between the soft gluons and the Coulomb gluons can therefore be a bit different from the case of $t\bar{t}$ production. This poses a question of how to properly factorize these contributions to all orders in perturbation theory, which was not addressed in the literature. Secondly, this process is closely related to the $e^+e^- \to t\bar{t}h$ process at future electron-positron colliders, which receives important QED corrections, especially in the threshold region \cite{Denner:2003zp}. The investigation of soft and Coulomb gluons in the $pp \to t\bar{t}h$ process can therefore be applied straightforwardly to the soft and Coulomb photons in the $e^+e^- \to t\bar{t}h$ process. Finally, while the small-$\beta$ limit is not significant for the total cross section, it could be important if one specifically wants to study certain kinematic configurations sensitive to the threshold region by, e.g., vetoing additional jets.

In this work, we are going to study the threshold region by applying a cut on the $\beta$ variable. While this is not a physical cut (since $\beta$ cannot be measured), it simplifies the theoretical considerations. We define the following quantity \cite{Moch:2008qy}
\begin{align}
\sigma_{\text{cut}}(s,m_t,m_h,\beta_{\text{cut}}) &= \sum_{ij}  \int^1_{\tau_{\text{min}}}d\tau \, \hat{\sigma}_{ij}(\tau,m_t,m_h,\mu_f) \, \theta(\beta_{\text{cut}}-\beta) \, \ff_{ij}(\tau,\mu_f) \, .
\label{eq:sigma_cut}
\end{align}
Apparently, for sufficiently small $\beta_{\text{cut}}$, $\sigma_{\text{cut}}$ should be well approximated by the leading power expression of the threshold expansion. It is also obvious that as $\beta_{\text{cut}} \to 1$, $\sigma_{\text{cut}}$ approaches the total cross section defined in Eq.~\eqref{eq:sigma}. The constrained cross section $\sigma_{\text{cut}}$ is the main object we are going to study in the rest of the paper. Before entering the technical details, we first investigate its leading order behavior against the variation of $\beta_{\text{cut}}$.

\begin{figure}[t!]
\centering
\includegraphics[width=0.6\textwidth]{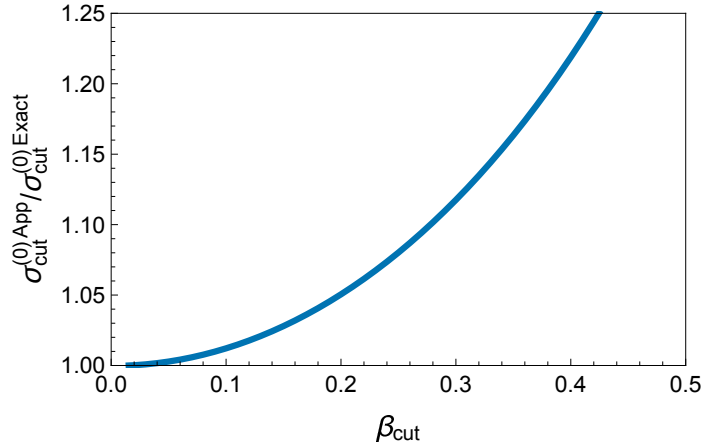}
\caption{Ratio between $\sigma^{(0)\text{App}}_{\text{cut}}$ and $\sigma^{(0)\text{Exact}}_{\text{cut}}$ as a function of $\beta_{\text{cut}}$ at \unit{13}{\TeV} LHC.}
\label{fig:sigma_cut_LO}
\end{figure}

We again choose to work with the \unit{13}{\TeV} LHC. We take the exact results and the approximate results for the leading order partonic cross sections $\hat{\sigma}^{(0)\text{Exact}}_{ij \to t\bar{t}h}$ and $\hat{\sigma}^{(0)\text{App}}_{ij \to t\bar{t}h}$, and plug them into Eq.~\eqref{eq:sigma_cut}. We denote the results as $\sigma^{(0)\text{Exact}}_{\text{cut}}$ and $\sigma^{(0)\text{App}}_{\text{cut}}$, and plot their ratio in Fig.~\ref{fig:sigma_cut_LO} as a function of $\beta_{\text{cut}}$. From the figure, we find that up to $\beta_{\text{cut}} \lesssim 0.3$, $\sigma^{(0)\text{App}}_{\text{cut}}$ provides a rather good approximation to the exact result. This fact should be kept in mind when we later combine the small-$\beta$ resummation and the fixed-order calculation.

At higher orders in perturbation theory, exchanges of Coulomb gluons and soft gluons lead to threshold-enhanced terms such as $1/\beta^n$ and $\ln^n\beta$. For small $\beta_{\text{cut}}$, these terms represent the dominant contributions to the constrained cross section $\sigma_{\text{cut}}$. The rest of the paper will be devoted to deriving a factorization formula for the partonic cross section in the threshold limit, and resumming these enhanced terms to all orders in perturbation theory.

\section{Factorization and resummation in the threshold limit}
\label{sec:factorization}

\subsection{Higher order QCD corrections in the threshold limit}

Beyond the Born level, the cross sections receive contributions from exchange of virtual gluons and emission of real gluons. We will investigate these contributions as a power expansion in $\beta$ in the threshold limit $\beta \to 0$. In this limit, there will be $\ln\beta$-enhanced terms and $1/\beta$-enhanced terms at higher orders in $\alpha_s$. Schematically, we are going to consider corrections of the form
\begin{multline}
\hat{\sigma}_{ij}^{\text{NLL}'} \sim \alpha_s^0 \Big\{ 1, \beta \Big\} + \alpha_s \Big\{ \ln^2\beta, \ln\beta, 1, \frac{1}{\beta}, \beta\ln^2\beta, \beta\ln\beta \Big\}
\\
+ \alpha^2_s \Big\{ \ln^4\beta, \ln^3\beta, \ln^2\beta, \frac{1}{\beta^2}, \frac{1}{\beta}, \frac{\ln^2\beta}{\beta}, \frac{\ln\beta}{\beta}, \beta\ln^4\beta, \beta\ln^3\beta \Big\} + \cdots \, .
\label{eq:NLLprimeExpansion}
\end{multline}
The collection of these terms are referred to as the improved next-to-leading logarithmic (NLL$'$) corrections. Note that due to the presence of two kinds of terms, one needs to insist on a consistent logarithmic counting for both of them, which we take as $\lambda \sim \alpha_s \sim \beta \sim 1/\ln\beta$. Using this counting, it can be seen that the NLL$'$ corrections include terms up to order $\lambda^1$. It is also clear from this counting that one needs to include formally $\mathcal{O}(\beta^1)$ next-to-leading power (NLP) terms besides the $\mathcal{O}(\beta^0)$ leading power (LP) ones in the power expansion. This greatly complicates the analysis of factorization, as will be clear below.

The behavior of higher order corrections in the threshold limit can be studied using the method of regions \cite{Beneke:1997zp, Jantzen:2011nz}. We work in the partonic center-of-mass frame where the momenta of the two incoming partons are given by
\begin{align}
p_1^\mu = \frac{\sqrt{\hat{s}}}{2} \, n^\mu \, , \quad p_2^\mu = \frac{\sqrt{\hat{s}}}{2} \, \bar{n}^\mu \, ,
\end{align}
where $n$ and $\bar{n}$ are two light-like vectors satisfying $n^2 = \bar{n}^2 = 0$ and $n \cdot \bar{n} = 2$. For a given momentum $k$, we perform the light-cone decomposition as
\begin{align}
k^\mu = \frac{k_+}{2} \, n^\mu + \frac{k_-}{2} \, \bar{n}^\mu + k_\perp^\mu \, ,
\end{align}
with $k_+ = \bar{n} \cdot k$ and $k_- = n \cdot k$. We identify the following momentum regions relevant to our problem:
\begin{equation}
\begin{array}{ll}
\text{hard}: & k^{\mu} \sim \sqrt{\hat{s}} \, ,
\\
\text{soft}: & k^{\mu} \sim \sqrt{\hat{s}} \, \beta \, ,
\\
\text{potential}: & k^0 \sim \sqrt{\hat{s}} \, \beta^2 \, , \quad \vec{k} \sim \sqrt{\hat{s}} \, \beta \, ,
\\
\text{ultrasoft}: & k^{\mu} \sim \sqrt{\hat{s}} \, \beta^2 \, ,
\\
\text{collinear}: & (k_+,k_-,k_{\perp}) \sim \sqrt{\hat{s}} \, (1,\beta^2,\beta) \, ,
\\
\text{anticollinear}: & (k_+,k_-,k_{\perp}) \sim \sqrt{\hat{s}} \, (\beta^2,1,\beta) \, .
\end{array}
\label{eq:ScalingRules2}
\end{equation}
These serve as the basis for constructing the effective field theoretic description of the process, and for deriving the factorization formula for the cross sections. At this point, it should be noted that there is a subtle difference between $t\bar{t}h$ production here and $t\bar{t}$ production discussed in \cite{Beneke:2009ye, Beneke:2009rj, Beneke:2010da, Beneke:2011mq, Piclum:2018ndt}. In $t\bar{t}$ production, the 3-momentum of the $t\bar{t}$ pair is of the ultrasoft scale $\sqrt{\hat{s}}\beta^2$. This means that the $t\bar{t}$ rest frame is formally equivalent to the partonic center-of-mass frame. On the other hand, in $t\bar{t}h$ production, the $t\bar{t}$ pair is recoiled by the Higgs boson and has a 3-momentum of the potential scale $\sqrt{\hat{s}}\beta$. Therefore, an ultrasoft mode in the partonic center-of-mass frame will become a potential mode in the $t\bar{t}$ rest frame. The impact of this difference on the factorization and resummation will be discussed in this section.

\subsection{Effective field theories}
\label{sec:eft}

In order to derive the factorization and resummation formulas in the threshold limit, it is useful to employ the language of effective field theories (EFTs). According to the momentum regions in Eq.~\eqref{eq:ScalingRules2}, the relevant EFTs are the soft-collinear effective theory (SCET) and the non-relativistic QCD (NRQCD).

SCET \cite{Bauer:2000ew, Bauer:2000yr, Bauer:2001yt, Beneke:2002ph, Beneke:2002ni} describes the interactions among collinear, anticollinear and ultrasoft modes. At leading power and next-to-leading power in $\beta$, the effective Lagrangians are given by
\begin{align}
\mathcal{L}_{\text{SCET}}^{0} &= \bar{\xi}_n \Big( i n \cdot D_n + g_s n \cdot A_{us} + i \slashed{D}_{n\perp} \frac{1}{i \bar{n} \cdot D_n} i \slashed{D}_{n\perp} \Big) \frac{\slashed{\bar{n}}}{2} \xi_n - \frac{1}{2} \Tr \Big\lbrace F^{\mu\nu}_n F_{\mu\nu}^n \Big\rbrace + ( n \leftrightarrow \bar{n} ) \nonumber
\\
&- \frac{1}{2} \Tr \Big\lbrace F^{\mu\nu}_{\text{us}} F_{\mu\nu}^{\text{us}}\Big\rbrace \, ,
\label{eq:L0SCET}
\\
\mathcal{L}_{\text{SCET}}^{1a} &= \bar{\xi}_n \Big(   x_{\perp}^{\mu} n^{\nu} \overline{W}_n g_s F^{\text{us}}_{\mu\nu} \overline{W}_{n}^{\dag} \Big)\frac{\slashed{\bar{n}}}{2} \xi_n + (n \leftrightarrow \bar{n}) \, ,
\\
\mathcal{L}_{\text{SCET}}^{1b} &= \Tr \bigg\lbrace n^{\mu} F_{\mu\nu}^n \overline{W}_n i\left[ x^{\rho}_{\perp} \bar{n}^{\rho} F^{\text{us}}_{\rho\sigma}, \overline{W}_n^{\dag} \left( iD^{\nu}_{n\perp} \overline{W}_n \right) \right] \overline{W}_n^{\dag} \bigg\rbrace - \Tr \Big\lbrace n^{\mu} F_{n}^{\mu\nu_{\perp}} \overline{W}_n \bar{n}^{\rho} F^{\text{us}}_{\rho\nu_{\perp}} \overline{W}_n^{\dag} \Big\rbrace \nonumber
\\
&+ (n \leftrightarrow \bar{n}) \, ,
\\
\mathcal{L}_{\text{SCET}}^{1c} &= \bar{\xi}_n i \slashed{D}_{n\perp} \overline{W}_n q_{\text{us}} + \text{h.c.} + (n \leftrightarrow \bar{n}) \, ,
\end{align}
where $\xi_{n}$ and $q_{\text{us}}$  denote the collinear and ultrasoft quark fields; $A_{n}$ and $A_{(\text{us})}$ represent the collinear (ultrasoft) gluon fields, with $F_{n(\text{us})}^{\mu\nu}$ their field strength tensors; $\overline{W}_n$ is the collinear Wilson line.

To describe the interactions among the potential, soft and ultrasoft modes, we employ the potential non-relativistic QCD (pNRQCD) \cite{Pineda:1997bj, Brambilla:1999xf, Beneke:1999zr, Beneke:1999qg}. The leading power and next-to-leading power effective Lagrangians can be written as \cite{Beneke:1999zr, Beneke:1999qg, Kniehl:2002br}
\begin{align}
\mathcal{L}_{\text{pNRQCD}}^{0}(x) &=\psi^{\dag} \left( i D_{\text{us}}^0 +\frac{\vec{\partial}^2}{2m_t} \right) \psi + \chi^{\dag} \left( i D_{\text{us}}^0 - \frac{\vec{\partial}^2 }{2m_t} \right) \chi \nonumber
\\
&- \int \mathrm{d}^3\vec{r} \, \psi^{\dag} T^{a} \psi \big( x^0,\vec{x}+\vec{r} \big) \left( \frac{\alpha_s}{r} \right) \chi^{\dag} T^{a}\chi \big(x^0,\vec{x}\big) \, ,
\label{eq:L0pNRQCD}
\\
\mathcal{L}_{\text{pNRQCD}}^{1a}(x) &= -\psi^{\dag}(x) \, g_s \, \vec{x} \cdot \vec{E}_{\text{us}}(x^0,\vec{0}) \, \psi(x) - \chi^{\dag}(x) \, g_s \, \vec{x} \cdot \vec{E}_{\text{us}}(x^0,\vec{0}) \, \chi(x) \, ,
\label{eq:L1apNRQCD}
\\
\mathcal{L}_{\text{pNRQCD}}^{1b}(x) &= - \int \mathrm{d}^3\vec{r} \, \psi^{\dag}T^{a} \psi \big( x^0,\vec{x}+\vec{r} \big) \frac{\alpha^2_s}{4\pi r} \Big[ a_1 + 2\beta_0 \, \ln \big( e^{\gamma_E} \mu r \big) \Big] \chi^{\dag} T^{a}\chi \big( x^0,\vec{x} \big) \, ,
\label{eq:L1bpNRQCD}
\end{align}
where $\psi$ and $\chi$ are Pauli spinor fields annihilating the top quark and creating the anti-top quark, respectively; $\vec{E}^{i}_{\text{us}}=F^{i0}_{\text{us}}$ are the chromoelectric components of the ultrasoft field strength tensor. The coefficient $a_1$ was calculated in \cite{Fischler:1977yf, Billoire:1979ih} and is given by $a_1 = 31 C_A/9 - 10 n_f/9$. The one-loop coefficient $\beta_0$ of the QCD $\beta$-function is given in the Appendix.
Note that in the pNRQCD power counting, $\alpha_s$ and $r$ in Eq.~\eqref{eq:L0pNRQCD} and \eqref{eq:L1bpNRQCD} are considered as order $\beta$.
Worthy of particular attention is that in Refs.~\cite{Beneke:1999zr, Beneke:1999qg, Kniehl:2002br}, the pNRQCD Lagrangian is derived in the rest frame of the quarkonium, where the heavy quark pair is recoiled by ultrasoft momenta.
This is different from our case of $t\bar{t}h$ production, where the top quark pair is recoiled by the Higgs boson.
However, since the LO and NLO potentials in Eqs.~\eqref{eq:L0pNRQCD}-\eqref{eq:L1bpNRQCD} only involve the relative momentum between the heavy quark and anti-quark, in our case the LP and NLP Lagrangians take the same form.
It should be stressed that this fact is not expected to hold beyond NLP.
For example, as shown in Appendix~\ref{app:nnlp}, new structures depending on the recoil momentum appear in the NNLP Lagrangian.

To derive the factorization formula, we first match the QCD amplitudes onto an effective Hamiltonian constructed out of the SCET and pNRQCD fields. Generically, we write
\begin{equation}
\mathcal{H} \equiv \mathcal{H}_{\text{LP}} + \mathcal{H}_{\text{NLP}} +\cdots \equiv \sum_{I,m} C_{\text{LP}}^{Im} O_{\text{LP}}^{Im} + \sum_{I,m} C_{\text{NLP}}^{Im} O_{\text{NLP}}^{Im} + \cdots \, ,
\label{eq:Heff}
\end{equation}
where $I$ labels different color structures and $m$ for Lorentz structures, $O^{Im}_{\text{LP}}$ and $O^{Im}_{\text{NLP}}$ are leading power and next-to-leading power effective operators describing the scattering process under consideration, while $C^{Im}_{\text{LP}}$ and $C^{Im}_{\text{NLP}}$ are their Wilson coefficients arising from the hard region contributions. Note that the NLP effective Hamiltonian $\mathcal{H}_{\text{NLP}}$ actually does not contribute to the cross section at next-to-leading power. The reason is that such a contribution would be given by the interferences between $O^{Im}_{\text{LP}}$ and $O^{Im}_{\text{NLP}}$,
which vanish due to angular momentum conservation.

To obtain the Wilson coefficients $C^{Im}_{\text{LP}}$, we need to calculate on-shell scattering amplitudes in the limit $\sqrt{\hat{s}} = 2m_t + m_h$ using both QCD and the effective Hamiltonian. In this limit, the loop integrals in the effective theories are scaleless and vanish in dimensional regularization. Therefore, we only need to calculate the QCD amplitudes up to NLO. For this calculation we employ the program packages FeynArts \cite{Hahn:2000kx}, FeynCalc \cite{Mertig:1990an, Shtabovenko:2016sxi} and FIRE5 \cite{Smirnov:2014hma} to generate the amplitudes and perform the reduction to master integrals. The resulting master integrals can be evaluated to analytic expressions using Package-X \cite{Patel:2015tea, Patel:2016fam}. The interface connecting Package-X, FIRE5 and FeynCalc is provided by FeynHelpers \cite{Shtabovenko:2016whf}. We renormalize the top quark mass in the on-shell scheme, and the strong coupling $\alpha_s$ in the $\overline{\text{MS}}$ scheme. The Wilson coefficients can then be extracted after renormalizing the effective operators (which is equivalent to subtracting the infrared poles from the QCD amplitudes). For the purpose of this paper, we don't need the explicit forms of individual Wilson coefficients and effective operators, but the combinations of them entering the cross section for $t\bar{t}h$ production. These combinations will be given in the next subsection as the ``hard functions''.

\subsection{Factorization at leading and next-to-leading power}

The discussion in the last subsection tells us that for the cross section up to NLP, we only need to consider the amplitudes of $\mathcal{H}_{\text{LP}}$. At leading power, we use $\mathcal{H}_{\text{LP}}$ together with the LP Lagrangians $\mathcal{L}_{\text{SCET}}^{0}$ and $\mathcal{L}_{\text{pNRQCD}}^{0}$ to calculate the cross section
\begin{multline}
\hat{\sigma}_{ij}^{\text{LP}} = \frac{1}{2\hat{s}} \int d\Phi_t \, d\Phi_{\bar{t}} \, d\Phi_h \sum_X (2\pi)^4\delta^{(4)}(p_1+p_2-p_t-p_{\bar{t}}-p_h-p_X)
\\
\times \overline{\sum_{\text{pol}}} \, \overline{\sum_{\text{color}}} \braket{ij | \mathcal{H}_{\text{LP}}^\dagger | t\bar{t}hX} \braket{t\bar{t}hX | \mathcal{H}_{\text{LP}} | ij} \, ,
\label{eq:xs}
\end{multline}
where $ij=q\bar{q},gg$ labels the initial state, and $d\Phi_f$ represents phase-space integration over the momentum $p_f$ of the final state particle $f$.
In the LP Lagrangians \eqref{eq:L0SCET} and \eqref{eq:L0pNRQCD}, the interactions of the ultrasoft gluon with the collinear fields and heavy quark fields are encoded in the covariant derivative $D_{\text{us}}$. Such interactions can be removed by the decoupling transformations \cite{Bauer:2001yt, Beneke:2010da}
\begin{gather}
\xi_{n(\bar{n})}(x) \to S^q_{n(\bar{n})}(x) \xi_{n(\bar{n})}(x) \, , \quad A_{n(\bar{n})}(x) \to S^g_{n(\bar{n})}(x) A_{n(\bar{n})}(x) \, , \nonumber
\\
\psi(x) \to S_v(x) \psi(x) \, , \quad \chi(x) \to S_v(x) \chi(x) \, ,
\label{eq:decoupling}
\end{gather}
where $S_v(x)$ and $S^{q}_{n(\bar{n})}$ are ultrasoft Wilson lines in the fundamental representation along the directions implied by the subscripts, while $S^{g}_{n(\bar{n})}$ are ultrasoft Wilson lines in the adjoint representation.

After the decoupling, the factorization of the LP cross section follows along the same line of arguments as the $t\bar{t}$ production \cite{Beneke:2010da}. The only difference comes from the appearance of the Higgs momentum $p_h$. The factorization formula therefore reads
\begin{equation}
\hat{\sigma}_{ij}^{\text{LP}} = \frac{1}{2\hat{s}} \int d\Phi_h d\omega \, H^{J I}_{ij}(\mu) \, J^\alpha_{\text{LP}} \bigg( E_J-\frac{\omega}{2},\vec{p}_J \bigg) \, S^{\alpha I J}_{ij}(\omega,\mu) \, ,
\label{eq:fac}
\end{equation}
where
\begin{equation}
E_J = \sqrt{\hat{s}} - 2m_t - m_h - \frac{|\vec{p}_h|^2}{2m_h} \, , \quad \vec{p}_J = -\vec{p}_h \, .
\label{eq:pJ0}
\end{equation}
The potential function is given by
\begin{multline}
J^\alpha_{\text{LP}}(q^0-2m_t,\vec{q}) = \int d\Phi_{t} \, d\Phi_{\bar{t}} \, (2\pi)^4\delta^{(4)}(q - p_t - p_{\bar{t}}) 
\\
\times \sum_{\text{pol}} \sum_{\text{color}} P^\alpha_{\{a\}}\braket{0 | \chi_{a_2s_2}^\dagger\psi_{a_1s_1} | t\bar{t}} \braket{t\bar{t} | \psi_{a_3s_1}^\dagger\chi_{a_4s_2} | 0} \, ,
\label{eq:JLP}
\end{multline}
where $\{a\} = \{a_1,a_2,a_3,a_4\}$ and $P^\alpha_{\{a\}}$ are the projectors to the singlet-octet color states of the $t\bar{t}$ system, which are given by
\begin{equation}
P^{(1)}_{\{a\}} = \frac{1}{3} \delta_{a_1a_2} \delta_{a_3a_4} \, , \quad P^{(8)}_{\{a\}} = 2T^c_{a_1a_2}T^c_{a_4a_3} \, .
\end{equation}
Note that since the interactions in the LP Lagrangian $\mathcal{L}^0_{\text{pNRQCD}}$ are spin-independent, the two $\psi$ fields in the definition of the potential function share the same polarization index $s_1$, and similar for the two $\chi$ fields. It should be stressed that the potential function here is different from that in $t\bar{t}$ production, due to the presence of the recoil momentum $\vec{p}_h$.

The soft function in Eq.~\eqref{eq:fac} is defined as
\begin{multline}
S^{\alpha I J}_{ij}(\omega,\mu) = \sum_X \delta(\omega - 2E_X) P^\alpha_{\{a\}} \mathcal{C}^{I*}_{ij,\{b\}} \mathcal{C}^J_{ij,\{c\}}
\\
\times \braket{0 | S^\dagger_{v,a_2b_4} S_{v,b_3a_1} S^{j}_{\bar{n},b_2d_2} S^{i\dagger}_{n,d_1b_1}| X} \braket{X | S^i_{n,c_1d_1} S^{j\dagger}_{\bar{n},d_2c_2} S^\dagger_{v,a_3c_3} S_{v,c_4a_4} | 0}  \, ,
\end{multline}
where the color basis for the $q\bar{q}$ channel is given by
\begin{equation}
\mathcal{C}^{(1)}_{q\bar{q},\{a\}} = \frac{1}{3} \delta_{a_2a_1} \delta_{a_3a_4} \, , \quad
\mathcal{C}^{(2)}_{q\bar{q},\{a\}} = \frac{1}{\sqrt{2}} T^c_{a_2a_1} T^c_{a_3a_4} \, ,
\end{equation}
and that for the $gg$ channel is
\begin{equation}
\mathcal{C}^{(1)}_{gg,\{a\}} = \frac{1}{2\sqrt{6}} \delta_{a_2a_1} \delta_{a_3a_4} \, , \quad
\mathcal{C}^{(2)}_{gg,\{a\}} = \frac{1}{2\sqrt{3}} F^c_{a_2a_1} T^c_{a_3a_4} \, , \quad
\mathcal{C}^{(3)}_{gg,\{a\}} = \frac{1}{2} \sqrt{\frac{3}{5}} D^c_{a_2a_1}T^c_{a_3a_4} \, ,
\end{equation}
where $F^c_{ab} = if^{acb}$ and $D^c_{ab} = d^{abc}$. This soft function is the same as that for $t\bar{t}$ production in \cite{Beneke:2010da}, since it does not feel the presence of the recoil momentum. It is diagonal in the color basis we have chosen. In practice, it is more convenient to quote its Laplace transform
\begin{equation}
\tilde{s}^{\alpha I J}_{ij}(L,\mu) = \int^{\infty}_{0} d\omega \, \exp \bigg( -\frac{
N\omega}{2m_t+m_h} \bigg) \, S^{\alpha I J}_{ij}(\omega,\mu) \, ,
\label{eq:laplace}
\end{equation} 
where
\begin{equation}
L = 2 \ln \frac{\mu N e^{\gamma_E}}{2m_t + m_h} \, .
\end{equation}
Up to the NLO, the result reads \cite{Beneke:2010da}
\begin{equation}
\tilde{s}^{\alpha II}_{ij}(L,\mu) = 1 + \frac{\alpha_s}{4\pi} \left[ (C_i + C_j) \left( L^2 + \frac{\pi^2}{6} \right) + 2C_\alpha (L+2) \right] ,
\end{equation}
where $C_i = C_F$ for $i = q,\bar{q}$, $C_i = C_A$ for $i = g$, $C_\alpha = 0$ for $\alpha = (1)$ (singlet), and $C_\alpha = C_A$ for $\alpha = (8)$ (octet). Note that the soft function actually does not depend on the index $I$.

Finally, the hard function $H^{I J}_{ij}$ is defined as the product of LP Wilson coefficients $C^{Im*}_{\text{LP}} C^{Jm}_{\text{LP}}$ projected onto the $S$-wave spin structure determined by the potential function. In general, the hard function is not diagonal. However, since the soft function is diagonal, only the diagonal entries of the hard function contribute to the cross section. Furthermore, since the soft function $\tilde{s}^{\alpha I I}_{ij}$ does not depend on the index $I$, we can project the diagonal entries of the hard function onto the singlet-octet basis as
\begin{equation}
H^{(1)}_{ij} = H^{11}_{ij} \, , \quad H^{(8)}_{q\bar{q}} = H^{22}_{q\bar{q}} \, , \quad H^{(8)}_{gg} = H^{22}_{gg} + H^{33}_{gg} \, .
\end{equation} 
We have calculated these entries $H^\alpha_{ij}$ at NLO explicitly using the method described in the last subsection, and write the result as
\begin{multline}
H^{\alpha}_{ij} = \frac{4\pi^2 \alpha_s^2 m_t^2}{v^2 m_h^2 (m_h+2 m_t)^4} H^{\text{LO},\alpha}_{ij}
\\
\times \bigg[ 1 + \frac{\alpha_s}{4\pi} \left[ -4(C_i + C_j) L_H^2 + \left( 4\beta_0 + \gamma_{ij,0}^{H,\alpha} \right) L_H + H^{\text{NLO},\alpha}_{ij} \right] \bigg] \, ,
\end{multline}
where $L_H = \ln(\mu/(2m_t+m_h))$, and the LO coefficients are given by
\begin{align}
H^{\text{LO},(1)}_{q\bar{q}} &= 0 \, , \nonumber
\\
H^{\text{LO},(8)}_{q\bar{q}} &= \frac{32}{9} \, , \nonumber
\\
H^{\text{LO},(1)}_{gg} &= \frac{C_F (m_h^2-4m_t^2)^2}{64m_t^4} \, , \nonumber
\\
H^{\text{LO},(8)}_{gg} &= -2 H^{\text{LO},(1)}_{gg} + \frac{C_A^2C_F (m_h^4-4m_h^2m_t^2+8m_t^4)}{64m_t^4} \, .
\end{align}
The one-loop hard anomalous dimensions $\gamma_{ij,0}^{H,\alpha}$ are given by
\begin{equation}
\gamma^{H,(8)}_{q\bar{q},0} = - 12 C_F - 4 C_A  \, , \quad \gamma^{H,(1)}_{gg,0} = -4 \beta_0 \, , \quad \gamma^{H,(8)}_{gg,0} = -4 \beta_0 - 4 C_A \, .
\label{eq:gammaH}
\end{equation}
The analytic expressions for the NLO coefficients $H^{\text{NLO},\alpha}_{ij}$ are too tedious to be shown. To get some impression of their sizes, we quote the numeric values here with $m_t=\unit{173.5}{\GeV}$ and $m_h=\unit{125.09}{\GeV}$:
\begin{equation}
H^{\text{NLO},(8)}_{q\bar{q}} \approx 4.93 \, , \quad H^{\text{NLO},(1)}_{gg} \approx 37.0 \, , \quad H^{\text{NLO},(8)}_{gg} \approx 23.8 \, .
\end{equation}

To achieve NLL$'$ accuracy for the resummation, we need to further consider next-to-leading power corrections to the cross section. These amount to contributions from the NLP interactions in $\mathcal{L}_{\text{SCET}}^{1a,1b,1c}$ and $\mathcal{L}_{\text{pNRQCD}}^{1a,1b}$ to the squared-amplitude in Eq.~\eqref{eq:xs}. In analogy to the arguments in \cite{Beneke:2009ye, Beneke:2010da} for $t\bar{t}$ production, it can be shown that single insertions of $\mathcal{L}_{\text{SCET}}^{1a,1b}$ give vanishing results due to angular momentum conservation, while $\mathcal{L}_{\text{SCET}}^{1c}$ does not contribute due to baryon number conservation. The terms in $\mathcal{L}_{\text{pNRQCD}}^{1b}$ involve subleading potentials between the top and anti-top quarks. These contributions can be incorporated by upgrading the potential function $J^\alpha(q)$ to the NLO, which we will discuss in the next subsection.
Finally, we need to consider the corrections induced by $\mathcal{L}_{\text{pNRQCD}}^{1a}$.

The terms in $\mathcal{L}_{\text{pNRQCD}}^{1a}$ involve extra interactions between ultrasoft gluons and potential modes, which are not removed by the decoupling transform \eqref{eq:decoupling}.
In \cite{Beneke:2010da}, it was proved that these interactions do not contribute to the $t\bar{t}$ cross section at NLP. However, the arguments there rely on the fact that the partonic center-of-mass frame and the $t\bar{t}$ rest frame are the same, and therefore the potential function does not depend on an external 3-momentum. However, for $t\bar{t}h$ production, the recoil momentum $\vec{p}_h$ from the extra Higgs boson spoils the proof, and we need to reinvestigate the contributions from $\mathcal{L}_{\text{pNRQCD}}^{1a}$ here.

\begin{figure}[t!]
\centering
\includegraphics[scale=1.2]{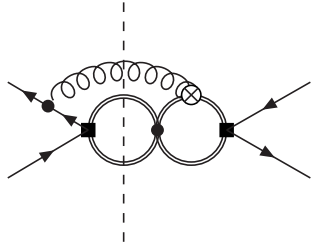}
\caption{\label{fig:example_1a}A sample Feynman diagram contributing to $\hat{\sigma}_{ij}^{1a}$. The black dots represent vertices from $\mathcal{L}_{\text{SCET}}^{0}$ and $\mathcal{L}_{\text{pNRQCD}}^{0}$. The black squares denote insertions of $\mathcal{H}_{\text{LP}}$. The circle with a cross stands for an insertion of $\mathcal{L}^{1a}_{\text{pNRQCD}}$.}
\end{figure} 

We begin with an explicit diagram depicted in Fig.~\ref{fig:example_1a}. Its contribution to the partonic cross section in the threshold limit can be written as (up to overall factors due to coupling constants, color factors, Wilson coefficients, etc.)
\begin{equation}
\Delta\hat{\sigma} \propto \int d\Phi_h \, \mathcal{A}(\vec{p}_h) \, ,
\label{eq:Delta_sigma}
\end{equation}
where $\mathcal{A}(\vec{p}_h)$ is given by
\begin{equation}
\mathcal{A}(\vec{p}_h) = \int d\Phi_t d\Phi_{\bar{t}} d\Phi_g \,
\delta(E_J - E_t - E_{\bar{t}} - p_g^0) \,
\delta^{(3)}(\vec{p}_h+\vec{p}_t+\vec{p}_{\bar{t}})
\int \frac{d^4k}{(2\pi)^4} \mathcal{M} \, ,
\end{equation}
where $E_J$ is defined in Eq.~\eqref{eq:pJ0},
\begin{equation}
E_t = \frac{|\vec{p}_t|^2}{2m_t} \, , \quad E_{\bar{t}} = \frac{|\vec{p}_{\bar{t}}|^2}{2m_t} \, ,
\end{equation}	
and
\begin{multline}
\mathcal{M} = \frac{1}{-|\vec{k}-\vec{p}_t|^2} \,
\frac{1}{k^0-p_g^0-|\vec{k}|^2/(2m_t)} \,
\frac{1}{E_t+E_{\bar{t}}+p_g^0-k^0-|\vec{k}-\vec{p}_t-\vec{p}_{\bar{t}}|^2/(2m_t)}
\\
\times \left( \frac{\partial}{\partial\vec{k}} \frac{1}{k^0-|\vec{k}|^2/(2m_t)} \right) \cdot \frac{(v \cdot n) \, \vec{p}_g - (v \cdot p_g) \, \vec{n}}{n \cdot p_g} \, .
\end{multline}
Note that we have suppressed the imaginary part $+i\varepsilon$ in the propagators.
We now observe that the last factor in the above expression does not depend on $\vec{k}$, $\vec{p}_{t}$, $\vec{p}_{\bar{t}}$ and $\vec{p}_h$, while the other factors do not depend on $\vec{n}$. Together with the fact that $\vec{v} = \vec{0}$, we can conclude that after integrating over $k$, $\vec{p}_t$, $\vec{p}_{\bar{t}}$ and $\vec{p}_g$, the function $\mathcal{A}(\vec{p}_h)$ must be proportional to $\vec{n} \cdot \vec{p}_h$ (multiplied by a function of $|\vec{p}_h|^2$ and other scalar quantities). As a result, after performing the integration over $\vec{p}_h$ as in Eq.~\eqref{eq:Delta_sigma}, the contribution of this diagram to the partonic total cross section must vanish.

The argument above can be generalized to all contributions from a single insertion of $\mathcal{L}_{\text{pNRQCD}}^{1a}$ in a more formal way.
The cross section induced by $\mathcal{L}_{\text{pNRQCD}}^{1a}$ can be written as
\begin{multline}
\hat{\sigma}_{ij}^{1a} = \frac{1}{2\hat{s}} \int d\Phi_t \, d\Phi_{\bar{t}} \, d\Phi_h \sum_X (2\pi)^4\delta^{(4)}(p_1+p_2-p_t-p_{\bar{t}}-p_h-p_X)
\\
\times \overline{\sum_{\text{pol}}} \, \overline{\sum_{\text{color}}}
\int d^4z \braket{ij | \mathcal{H}_{\text{LP}}^\dagger(0) | t\bar{t}hX} \braket{t\bar{t}hX | \mathrm{T} \big[ i\mathcal{L}_{\text{pNRQCD}}^{1a}(z) \, \mathcal{H}_{\text{LP}}(0) \big]| ij} + \text{h.c.} \, ,
\label{eq:xs1a}
\end{multline}
where $\mathrm{T}$ denotes time-ordered product.
We can perform the usual decoupling transforms \eqref{eq:decoupling} to remove the leading power interaction between ultrasoft and potential modes. The remaining interaction is of the $\vec{x} \cdot \vec{E}_{\text{us}}$ form from $\mathcal{L}_{\text{pNRQCD}}^{1a}$. As a result, we can write the cross section as
\begin{align}
\hat{\sigma}_{ij}^{\text{1a}} &= \frac{1}{2\hat{s}} \int d\Phi_h d\omega \, H_{ij}(\mu) \int \frac{d^4k}{(2\pi)^4} J_{\text{1a}} \bigg( E_J-\frac{\omega}{2},\vec{p}_J,k \bigg) \int d^4z \, e^{-i k \cdot z} \, i\vec{z} \cdot \vec{S}^{1a}_{ij}(\omega,z^0,\mu) + \text{h.c.} \nonumber
\\
&= \frac{1}{2\hat{s}} \int d\Phi_h d\omega dz^0 \frac{dk^0}{2\pi} \, e^{-i k^0z^0} H_{ij}(\mu) \, \vec{j}_{1a} \bigg( E_J-\frac{\omega}{2},\vec{p}_J,k^0 \bigg) \cdot \vec{S}^{1a}_{ij}(\omega,z^0,\mu) + \text{h.c.} \, ,
\label{eq:fac1a}
\end{align}
where we have suppressed all color indices for simplicity, while $E_J$ and $\vec{p}_J$ are given in Eq.~\eqref{eq:pJ0}. The subleading potential function and soft function are defined as
\begin{align}
J_{\text{1a}}(E_q,\vec{q},k) &= -\int d\Phi_{t} \, d\Phi_{\bar{t}} \, (2\pi)^4\delta^{(4)}(q - p_t - p_{\bar{t}}) \nonumber
\\
&\quad \times \int d^4x \, e^{i k \cdot x} \braket{0 | \chi_{s_2}^\dagger\psi_{s_1}(0) | t\bar{t}} \braket{t\bar{t} | \mathrm{T} \big[ \big( \psi^\dagger(x)\psi(x) + \chi^\dagger(x) \chi(x) \big) \, \psi_{s_1}^\dagger\chi_{s_2}(0) \big] | 0} \, , \nonumber
\\
\vec{S}^{1a}_{ij}(\omega,z^0,\mu) &= g_s \sum_X \delta(\omega - 2E_X) \nonumber
\\
&\quad \times
\braket{0 | S^\dagger_{v} S_{v} S^{j}_{\bar{n}} S^{i\dagger}_{n} (0) | X}\braket{X | \mathrm{T} \big[ S_{v} \vec{E}_{\text{us}} S^\dagger_v (z^0,\vec{0})  \, S^i_{n} S^{j\dagger}_{\bar{n}} S^\dagger_{v} S_{v} (0) \big] | 0}  \, ,
\end{align}
with $E_q = q^0 - 2m_t$, and
\begin{equation}
\vec{j}_{1a}(E_q,\vec{q},k^0) = -\int d^3\vec{k} \, \delta^{(3)}(\vec{k}) \,
\frac{\partial}{\partial \vec{k}} J_{\text{1a}}(E_q,\vec{q},k) \, ,
\end{equation}
where again we have ignored all color structures which are not important for the arguments here. Note that $\vec{j}_{1a}(E_q,\vec{q},k^0)$ must be proportional to $\vec{q}$ since this is the only 3-vector it can depend on.

For $t\bar{t}$ production, there is no recoil momentum and $\vec{p}_J = \vec{0}$. Therefore in the integrand for the cross section one has $\vec{j}_{1a}(E,\vec{0},k^0) = 0$, and one can conclude that the contribution from $\mathcal{L}_{\text{pNRQCD}}^{1a}$ to the cross section vanishes. This is essentially the argument in \cite{Beneke:2010da}. For $t\bar{t}h$ production, due to the presence of a recoil momentum, $\vec{j}_{1a}$ can depend on $\vec{p}_J = -\vec{p}_h$, and is not zero in general. However, note that the whole integrand in Eq.~\eqref{eq:fac1a} is an odd function of $\vec{p}_h$. Consequently, after integrating over the phase space $d\Phi_h$ of the Higgs boson, the contribution still vanishes. Therefore, we arrive at the same conclusion as in the $t\bar{t}$ case that the only NLP contribution to the total cross section comes from $\mathcal{L}_{\text{pNRQCD}}^{1b}$. We emphasize that this fact only holds at the level of total cross section, and extra corrections may be present if one does not integrate over the momentum of the Higgs boson.

In summary, up to the next-to-leading power, the cross section can be factorized as 
\begin{equation}
\hat{\sigma}_{ij} = \sum_{\alpha} \frac{1}{2\hat{s}} \int d\Phi_h d\omega \, H^\alpha_{ij}(\mu) \, J^\alpha\bigg( E_J-\frac{\omega}{2},\vec{p}_J \bigg) \, S^{\alpha}_{ij}(\omega,\mu) \, ,
\label{eq:fac2}
\end{equation}
where the hard function $H$ and soft function $S$ only receive leading power contributions. The potential function $J^\alpha(q)$ contains both LP and NLP contributions, which we present in the next subsection. Note that this simple form of the factorization formula is not expected to hold at higher powers in $\beta$, as we'll discuss in Appendix~\ref{app:nnlp}.

\subsection{The potential function with a recoil momentum}

As introduced in the last subsection, a non-trivial difference between $t\bar{t}h$ production and $t\bar{t}$ production is the dependence of the potential function $J^\alpha(E_q,\vec{q})$ on the recoil momentum $\vec{p}_h$ of the extra Higgs boson. Up to the NLO, we write the potential function as
\begin{equation}
J^\alpha(E_q,\vec{q}) = J^\alpha_{\text{LP}}(E_q,\vec{q}) + J^\alpha_{\text{NLP}}(E_q,\vec{q}) \, ,
\end{equation}
where the LP term is defined in Eq.~\eqref{eq:JLP}, and the NLP term is given by
\begin{multline}
J^\alpha_{\text{NLP}}(E_q,\vec{q}) = \int d\Phi_{t} \, d\Phi_{\bar{t}} \, (2\pi)^4\delta^{(4)}(q - p_t - p_{\bar{t}}) 
\\
\times \sum_{\text{pol}} \sum_{\text{color}} P^\alpha_{\{a\}} \int d^4x \braket{0 | \chi_{a_2s_2}^\dagger\psi_{a_1s_1}(0) | t\bar{t}} \braket{t\bar{t} | \mathrm{T} \big[ i\mathcal{L}^{1b}_{\text{pNRQCD}}(x) \, \psi_{a_3s_1}^\dagger\chi_{a_4s_2}(0) \big] | 0} + \text{h.c.} \, .
\label{eq:JNLP}
\end{multline}
In calculating the potential function, one needs to consider the interactions induced by the leading power Lagrangian $\mathcal{L}_{\text{pNRQCD}}^0$ to all orders in $\alpha_s$. In this way, one resums all terms of the form $(\alpha_s/\beta)^n$ and $\alpha_s(\alpha_s/\beta)^n$. According to \cite{Beneke:2010da}, the potential function can be related to the imaginary part of the pNRQCD Green function $G^\alpha(\vec{r}_1,\vec{r}_2;E_q,\vec{q})$ of the $t\bar{t}$ pair at origin:
\begin{equation}
J^\alpha(E_q,\vec{q}) = 2 \Imag G^\alpha(\vec{0}, \vec{0}; E_q, \vec{q}) \, ,
\label{eq:JE}
\end{equation}
where the Green function is defined as
\begin{equation}
G^\alpha(\vec{0}, \vec{0}; E_q, \vec{q}) = P^\alpha \int d^4z e^{i(q-2m_tv) \cdot z} \braket{ 0 | \mathrm{T} \big[ \chi^\dagger \psi (z) \, \psi^\dagger \chi (0) \big] | 0} \, ,
\end{equation}
where we have suppressed the color and Lorentz indices for simplicity.

For $t\bar{t}$ production at threshold, one needs the potential function with $E_q = E$ and $\vec{q} = \vec{0}$.
Up to the NLO, the result is given by \cite{Beneke:1999qg, Beneke:1999zr, Pineda:2006ri, Beneke:2011mq}
\begin{equation}
G^\alpha(\vec{0},\vec{0}; E, \vec{0}) = G^\alpha_0(\vec{0},\vec{0}; E) + G^\alpha_1(\vec{0},\vec{0}; E) \, ,
\end{equation}
where
\begin{align}
G^\alpha_0(\vec{0},\vec{0}; E) &= \frac{m_t^2}{4\pi} \bigg\lbrace - \sqrt{\frac{-E}{m_t}} + \frac{\alpha_s D_{\alpha}}{2} \Big[ -2L_J + 2\psi(\lambda) + 2\gamma_{E} - 1 \Big] \bigg\rbrace \, , \nonumber
\\
G^\alpha_1(\vec{0},\vec{0}; E) &= -\frac{m_t^2 D_{\alpha} \alpha^2_s}{16\pi^2} \bigg\lbrace a_1 \Big[ L_J + (1-\lambda) \psi'(\lambda) - \psi(\lambda) - \gamma_E \Big] \nonumber
\\
&\hspace{-4em} + \beta_0 \Big[ L_J^2 + 2L_J \big( (1-\lambda) \psi'(\lambda) - \psi(\lambda) - \gamma_E \big) + 4 {}_4F_3(1,1,1,1;2,2,\lambda;1) \nonumber
\\
&\hspace{-4em} + (1-\lambda) \psi''(\lambda) - 2 (1-\lambda) \big( \psi(\lambda) + \gamma_E \big) \psi'(\lambda) - \frac{\pi^2}{6} - 3\psi'(\lambda) + \big( \psi(\lambda) + \gamma_E \big)^2 \Big] \bigg\rbrace \, .
\end{align}
Here
\begin{equation}
L_J = - \frac{1}{2} \ln \bigg( -\frac{4m_t E}{\mu^2} \bigg) \, , \quad \lambda = 1 + \frac{\alpha_s D_{\alpha}}{2\sqrt{-E/m_t}} \, ,
\end{equation}
and $D_{(1)}=-C_F$, $D_{(8)}=1/(2N_c)$. To account for the finite width effects, one may replace $E \to E + i\Gamma_t$ in the above formulas, where $\Gamma_t$ is the width of the top quark.

We would now like to relate the potential function with a recoil momentum $J^\alpha(E_q,\vec{q})$ to the zero-recoil one $J^\alpha(E,\vec{0})$ given above. From the perturbative point-of-view, we can write the potential function as a sum of diagrams with arbitrary numbers of insertion of $\mathcal{L}_{\text{pNRQCD}}^0$ and up to one insertion of $\mathcal{L}_{\text{pNRQCD}}^{1b}$:
\begin{align}
J^{\alpha}(E_q,\vec{q}) &= \int \frac{d^3\vec{q}_1}{(2\pi)^3}
\frac{i}{E_q - \frac{|\vec{q}|^2}{4m_t} - \frac{|\vec{q}_1|^2}{m_t} + i \epsilon} \nonumber
\\
&- \int \frac{d^3\vec{q}_1}{(2\pi)^3} \frac{d^3\vec{q}_2}{(2\pi)^3}
\frac{i}{E_q - \frac{|\vec{q}|^2}{4m_t} - \frac{|\vec{q}_1|^2}{m_t} + i \epsilon} \,
\frac{i}{E_q - \frac{|\vec{q}|^2}{4m_t} - \frac{|\vec{q}_2|^2}{m_t} + i \epsilon} \nonumber
\\
&\hspace{3em} \times \frac{4\pi i \alpha_s D_{\alpha}}{|\vec{q}_1-\vec{q}_2|^2}
\left\{ 1 + \frac{\alpha_s}{4\pi} \left[ a_1 - \beta_0 \ln \bigg( \frac{|\vec{q}_1-\vec{q}_2|^2}{\mu^2} \bigg) \right] \right\} + \cdots \, ,
\end{align}
where the ellipsis denotes more insertions of heavy quark propagators and LO potential terms.
Note that the perturbative expansion of $J^\alpha(E,\vec{0})$ is the same as the above if we identify $E = E_q - |\vec{q}|^2/(4m_t)$. This identity can also be seen if we evaluate the potential function \eqref{eq:JLP} and \eqref{eq:JNLP} in the $t\bar{t}$ rest frame.
We can therefore deduce the relation
\begin{equation}
J^\alpha(E_q,\vec{q}) = J^\alpha \bigg( E_q - \frac{|\vec{q}|^2}{4m_t}, \vec{0} \bigg) \, , 
\label{eq:Jrelation}
\end{equation}
from which we obtain the potential function we need.

\subsection{Resummation}
\label{sec:resummation}

After deriving the factorization formula \eqref{eq:fac2}, we will now perform the resummation of both $1/\beta$ and $\ln\beta$ enhanced corrections at the NLL$'$ accuracy. Schematically, the resummed result takes the form
\begin{align}
\hat{\sigma}_{ij} &\sim \hat{\sigma}_{ij}^{\text{Born}} \sum_{k} \left( \frac{\alpha_s}{\beta} \right)^k
\exp \left[ \underbrace{\ln\beta f_0(\alpha_s\ln\beta)}_{(\text{LL})} +
\underbrace{f_1(\alpha_s\ln\beta)}_{(\text{NLL,NLL$'$})} +
\underbrace{\alpha_s f_2(\alpha_s\ln\beta)}_{(\text{NNLL,NNLL$'$})} + \cdots \right] \nonumber
\\
&\times \big\{ 1 (\text{LL,NLL}); \alpha_s, \beta (\text{NLL$'$,NNLL}); \alpha_s^2, \alpha_s\beta, \beta^2 (\text{NNLL$'$,NNNLL}) \big\} \, ,
\end{align}
where the counting rule is in accordance with \cite{Beneke:2009rj, Piclum:2018ndt}. The $1/\beta$ terms in the above expression are contained in the potential function $J^\alpha(E_q,\vec{q})$ in the factorization formula. To resum the $\ln\beta$ terms, we need to evaluate the hard function $H^\alpha_{ij}(\mu)$ and the soft function $S^\alpha_{ij}(\omega,\mu)$ separately at the hard scale $\mu_h$ and the soft scale $\mu_s$, and then use their renormalization group equations (RGEs) to evolve them to the factorization scale $\mu_f$.

The Laplace-space soft function $\tilde{s}^\alpha_{ij}$ satisfies the RGE
\begin{equation}
\frac{d}{d\ln\mu} \tilde{s}_{ij}^{\alpha}(L,\mu) = \left[
2 (C_i + C_j) \gamma_{\text{cusp}}(\alpha_s) \ln \frac{\mu N e^{\gamma_E}}{2m_t+m_h}
- \gamma^{s,\alpha}(\alpha_s) \right] \tilde{s}^{\alpha}_{ij}(L,\mu) \, ,
\label{eq:sRGE}
\end{equation}
where the cusp anomalous dimension $\gamma_{\text{cusp}}(\alpha_s)$ is given by \cite{Korchemskaya:1992je}
\begin{equation}
\gamma_{\text{cusp}}(\alpha_s) = \frac{\alpha_s}{\pi} + \left( \frac{\alpha_s}{4\pi} \right)^2 \left[
\left( \frac{268}{9} - \frac{4\pi^2}{3} \right) C_A - \frac{80}{9} n_f T_F \right] + \mathcal{O}(\alpha_s^3) \, ,
\end{equation}
and the soft anomalous dimension $\gamma^{s,\alpha}(\alpha_s)$ is \cite{Beneke:2009rj}
\begin{equation}
\gamma^{s,\alpha}(\alpha_s) = -\frac{\alpha_s}{\pi} C_\alpha + \mathcal{O}(\alpha_s^2) \, .
\end{equation}
Here we recall that $C_\alpha = 0$ for $\alpha = (1)$ (singlet) and $C_\alpha = C_A$ for $\alpha = (8)$ (octet). 

The RGE for the hard function is given by
\begin{equation}
\frac{d}{d\ln\mu} H_{ij}^{\alpha}(\mu) = \left[ 2 (C_i + C_j) \gamma_{\text{cusp}}(\alpha_s) \ln\frac{2m_t+m_h}{\mu} + \gamma^{H,\alpha}_{ij}(\alpha_s) \right] H_{ij}^{\alpha}(\mu) \, ,
\label{eq:hRGE}
\end{equation}
where the hard anomalous dimension can be expanded as
\begin{equation}
\gamma^{H,\alpha}_{ij}(\alpha_s) = \sum_{n=0}^{\infty} \left( \frac{\alpha_s}{4\pi} \right) \gamma^{H,\alpha}_{ij,n} \, ,
\end{equation}
with the one-loop coefficients given in Eq.~\eqref{eq:gammaH}.

The method to solve the RGEs \eqref{eq:sRGE} and \eqref{eq:hRGE} is standard \cite{Becher:2006mr, Becher:2007ty}. Plugging the result back to the factorization formula \eqref{eq:fac2}, we obtain the resummed cross section as
\begin{multline}
\hat{\sigma}_{ij}^{\text{resummed}}(\mu_f) = \sum_\alpha \frac{1}{2\hat{s}} \int d\Phi_h d\omega \, U_{ij}^\alpha(\mu_f,\mu_h,\mu_s) \, H^\alpha_{ij}(\mu_h) \, J^\alpha \bigg( E_J-\frac{\omega}{2}, \vec{p}_J \bigg)
\\
\tilde{s}^\alpha_{ij}(-\partial_\eta, \mu_s) \frac{e^{-2\gamma_E\eta}}{\Gamma(2\eta)} \left[ \frac{1}{\omega} \left( \frac{\omega}{\mu_s} \right)^{2\eta} \right]_* \, ,
\label{eq:resummed}
\end{multline}
where $\eta= A_{ij}^{\text{cusp}}(\mu_s,\mu_f)$ with
\begin{equation}
A_{ij}^{\text{cusp}}(\mu_a,\mu_b) = - \int^{\alpha_s(\mu_b)}_{\alpha_s(\mu_a)} d\alpha_s \frac{(C_i+C_j)\gamma_{\text{cusp}}(\alpha_s)}{\beta(\alpha_s)} \, .
\end{equation}
The explicit expressions for the QCD $\beta$-function and the evolution factor $U_{ij}^\alpha(\mu_f,\mu_h,\mu_s)$ are collected in Appendix~\ref{app:coeff}.

An important validation of the resummed formula \eqref{eq:resummed} is to compare its fixed-order expansion to explicit calculations. We define the coefficients of the expansion up to the NLO according to
\begin{equation}
\hat{\sigma}_{ij}^{\text{Expansion}}(\mu_f) = \sum_\alpha \hat{\sigma}^{\alpha,(0)}_{ij} \left[ 1 + \frac{\alpha_s}{4\pi} c^{\alpha,(1)}_{ij} \right] .
\label{eq:sigma_expansion}
\end{equation}
where
\begin{align}
\hat{\sigma}^{\alpha,(0)}_{ij} &= \frac{\beta^4 m_t^3 \sqrt{m_h(m_h+2m_t)}}{64\pi^2} H^{\text{LO},\alpha}_{ij} \, , \nonumber
\\
c^{\alpha,(1)}_{ij} &= (C_i+C_j) \left( 4L_S^2 -12 L_S + 14 - \frac{\pi^2}{2} \right) + C_\alpha (-4L_S + 10) \nonumber
\\
&-\frac{32\sqrt{2} \pi m_t D_\alpha}{3\sqrt{m_t(m_h+2m_t)}} \frac{1}{\beta} + H^{\alpha,(1)}_{ij} \, ,
\label{eq:coeff_expansion}
\end{align}
with $L_S = \ln(\beta^2(m_h+2m_t)/\mu)$. The above results contain the leading terms at LO and NLO in the threshold limit, and can be compared to explicit calculations performed in \cite{Beenakker:2002nc}. We have found complete agreement between the two results.

\section{Numeric results}
\label{sec:numerics}

\subsection{Matching to the NLO}

In this section, we present some numeric results based on our resummation formula Eq.~\eqref{eq:resummed}. Since the resummation formula contains only up to NLP terms, it is necessary to include higher power contributions whenever possible. For this reason we need to match the resummation formula to fixed-order calculations in order to extend the validity of our predictions beyond the threshold limit.
To take into account the higher power effects contained in the exact LO and NLO results, we use the following matching formula
\begin{equation}
\hat{\sigma}^{\text{matched}}_{ij}(\mu_f) = \sum_\alpha
\frac{\hat{\sigma}_{ij,\alpha}^{(0),\text{Exact}}}{\hat{\sigma}_{ij,\alpha}^{(0),\text{App}}}
\left[ \hat{\sigma}^{\text{resummed}}_{ij,\alpha}(\mu_f) - \hat{\sigma}^{\text{Expansion}}_{ij,\alpha}(\mu_f) \right] \theta(\beta_{\text{cut}}-\beta) + \hat{\sigma}^{\text{Exact}}_{ij}(\mu_f) \, ,
\label{eq:matching}
\end{equation}
where we have suppressed the dependence of the cross sections on other parameters for simplicity.
We have rescaled the resummed formula by the ratio between the exact and approximate LO results in each color channel. The resummed cross sections are calculated using Eq.~\eqref{eq:resummed} with NLL$'$ accuracy. The exact LO and NLO cross sections are calculated using Madgraph5\_aMC@NLO. The accuracy of the matched cross section is then denoted as NLL$'$+NLO.
Convoluting this matched partonic cross section with the parton luminosity function, we define the matched hadronic cross section 
\begin{align}
\sigma^{\text{matched}}(s,m_t,m_h,\beta_{\text{cut}}) &= \sum_{ij}  \int^1_{\tau_{\text{min}}} d\tau \, \hat{\sigma}_{ij}^{\text{matched}}(\tau,m_t,m_h,\mu_f) \, \ff_{ij}(\tau,\mu_f) \, .
\end{align}
This will be the main quantity for which we present numeric results later.

\subsection{The choice of the scales}

The resummed formula \eqref{eq:resummed} involves a set of unphysical scales, which are the hard scale $\mu_h$, the soft scale $\mu_s$ and the factorization scale $\mu_f$. In additional, although the potential function $J^\alpha(q)$ is formally independent of a renormalization scale, its finite order truncation has a residue scale-dependence. We denote this scale as $\mu_J$. These scales must be chosen appropriately to improve the convergence of the perturbation theory. This subsection is devoted to this issue.

From the explicit form of the hard function, one observes the appearance of the logarithms $\ln^n((2m_t+m_h)/\mu)$ at higher orders in perturbation theory. It is therefore natural to choose $\mu_h \sim 2m_t + m_h$ to make these logarithmic corrections under control. For the potential function, one observes logarithms of the form $\ln(4m_t\beta^2/\mu^2)$. However, the natural choice $\mu_J \sim 2m_t\beta$ requires an infrared cut-off, since the variable $\beta$ is integrated over from 0. We therefore choose
\begin{equation}
\mu_J^{\text{default}} = \max (2m_t\beta, \mu_J^{\text{cut}}) \, ,
\end{equation}
where $\mu_J^{\text{cut}}$ should be much larger than $\Lambda_{\text{QCD}}$. We follow the choice of \cite{Beneke:2010da} where $\mu_J^{\text{cut}}$ is set to be the solution to the equation $\mu_J^{\text{cut}} = C_F m_t \alpha_s(\mu_J^{\text{cut}})$. We solve this equation numerically and find $\mu_J^{\text{cut}} \approx \unit{32}{\GeV}$ which we take as the default value.
The factorization scale $\mu_f$ should be chosen to be appropriate in fixed-order calculations, since it enters the matching formula \eqref{eq:matching}. For that we take the conventional choice $\mu_f^{\text{default}} = m_t + m_h/2$.

\begin{figure}[t!]
\centering
\includegraphics[scale=0.82]{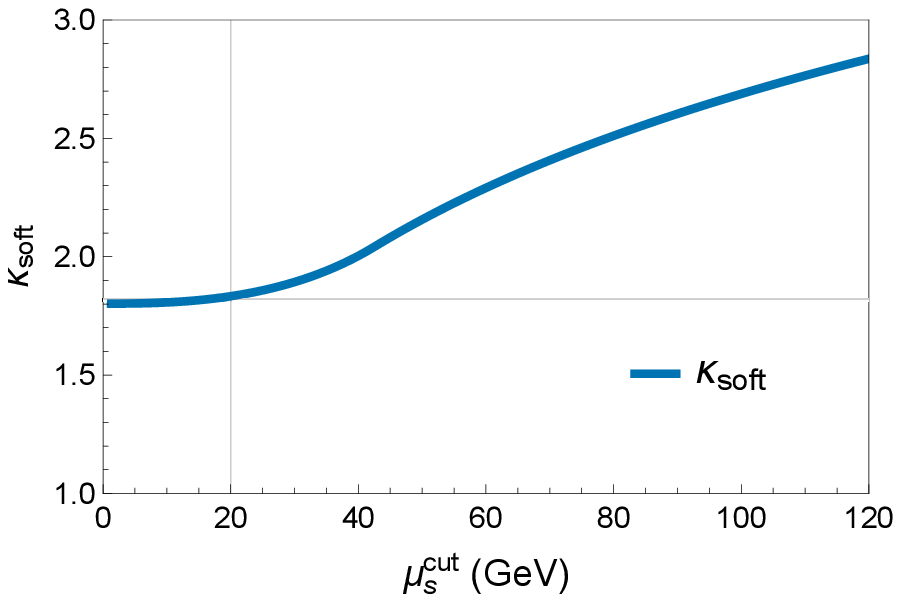}
\includegraphics[scale=0.82]{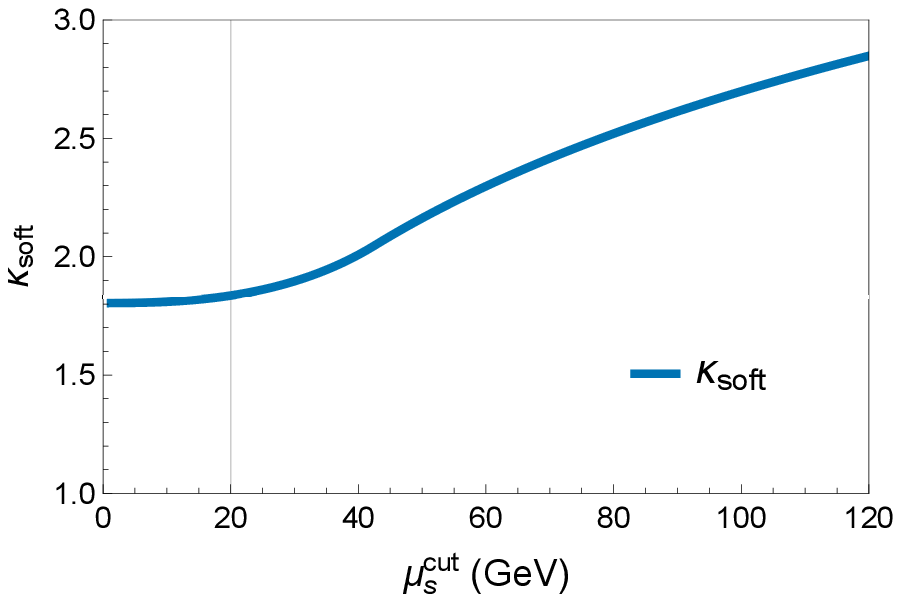}
\caption{\label{fig:kappa_SoftFinder}The soft ratio $\kappa_{\text{soft}}$ as a function of $\mu_s^{\text{cut}}$ at the $\unit{13}{\TeV}$ (left) and $\unit{14}{\TeV}$ (right) LHC.}
\end{figure}

The choice of the soft scale $\mu_s$ is more subtle. In the literature there are two kinds of methods for that purpose. One is to choose $\mu_s$ in the Laplace moment space, e.g., $\mu_s \sim (2m_t+m_h)/N$ where $N$ is the moment variable entering the Laplace transform Eq.~\eqref{eq:laplace}. Another is to choose $\mu_s$ in the momentum space, where $\mu_s \sim (2m_t+m_h)\beta^2$. We will use the latter method, for which we need to impose an infrared cut-off in order to avoid the Landau pole when integrating over $\beta$. Namely we have
\begin{equation}
\mu_s^{\text{default}} = \max ((2m_t+m_h)\beta^2, \mu_s^{\text{cut}}) \, ,
\end{equation}
The value of $\mu_s^{\text{cut}}$ should be much larger than $\Lambda_{\text{QCD}}$, but should not be too large since that will reintroduce large logarithms into the soft function. To study the effect of varying $\mu_s^{\text{cut}}$, we define the following quantity
\begin{align}
\sigma_{\text{cut}}^{\text{soft}}(s,m_t,m_h,\beta_{\text{cut}},\mu_f) &= \sum_{ij}  \int^1_{\tau_{\text{min}}} d\tau \, \hat{\sigma}_{ij}^{\text{soft}}(\tau,m_t,m_h,\mu_f) \, \theta(\beta_{\text{cut}}-\beta) \, \ff_{ij}(\tau,\mu_f) \, ,
\label{eq:sigma_cut_soft}
\end{align}
where $\hat{\sigma}_{ij}^{\text{soft}}$ is obtained from the NLO expansion Eq.~\eqref{eq:sigma_expansion} by removing the $1/\beta$ term and the one-loop hard function from the coefficient $c_{ij}^{\alpha,(1)}$ in Eq.~\eqref{eq:coeff_expansion}. The quantity $\sigma_{\text{cut}}^{\text{soft}}$ represents the contributions from the soft logarithms to the cross section in the threshold limit. We further define the ratio $\kappa_{\text{soft}}$ as
\begin{equation}
\kappa_{\text{soft}}(\mu_s^{\text{cut}}) = \left. \frac{\sigma_{\text{cut}}^{\text{soft}}(s,m_t,m_h,\beta_{\text{cut}},\mu_f)}{\sigma_{\text{cut}}^{(0)\text{App}}(s,m_t,m_h,\beta_{\text{cut}})} \right|_{\mu_f=\mu_s^{\text{default}}} \, ,
\end{equation}
where the denominator is simply the leading order part of the numerator. The above quantity represents the relative size of the soft corrections, and can be used to motivate an appropriate choice for $\mu_s^{\text{cut}}$. In Fig.~\ref{fig:kappa_SoftFinder}, we show the numeric values of $\kappa_{\text{soft}}$ as a function of $\mu_s^{\text{cut}}$. It can be seen that the size of the soft corrections stays stable when $\mu_s^{\text{cut}} < \unit{20}{\GeV}$, and increases dramatically when going beyond. Therefore, we choose the default value of $\mu_s^{\text{cut}}$ to be $\unit{20}{\GeV}$ in our numeric study.

\subsection{Results and discussions}

In this subsection, we present the numeric results for the total cross section at $\unit{13}{\TeV}$ and $\unit{14}{\TeV}$ LHC. For readers' convenience, we list here again the parameters we use: $m_t=\unit{173.5}{\GeV}$, $m_h=\unit{125.09}{\GeV}$ and $v=\unit{246.22}{\GeV}$. We have employed the MMHT2014 (N)LO PDFs \cite{Harland-Lang:2014zoa} with the corresponding $\alpha_s(m_Z)$.

\begin{figure}[t!]
\centering
\includegraphics[scale=0.83]{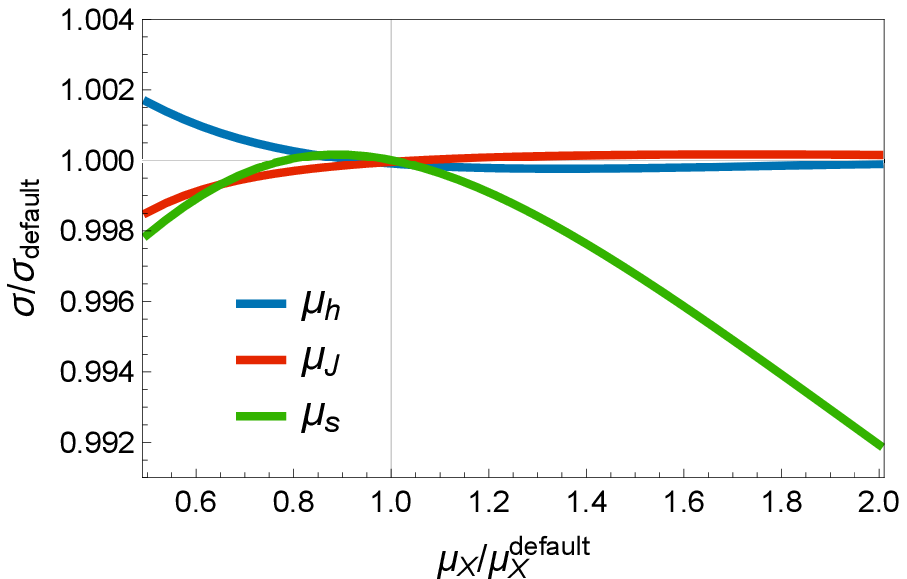}
\includegraphics[scale=0.80]{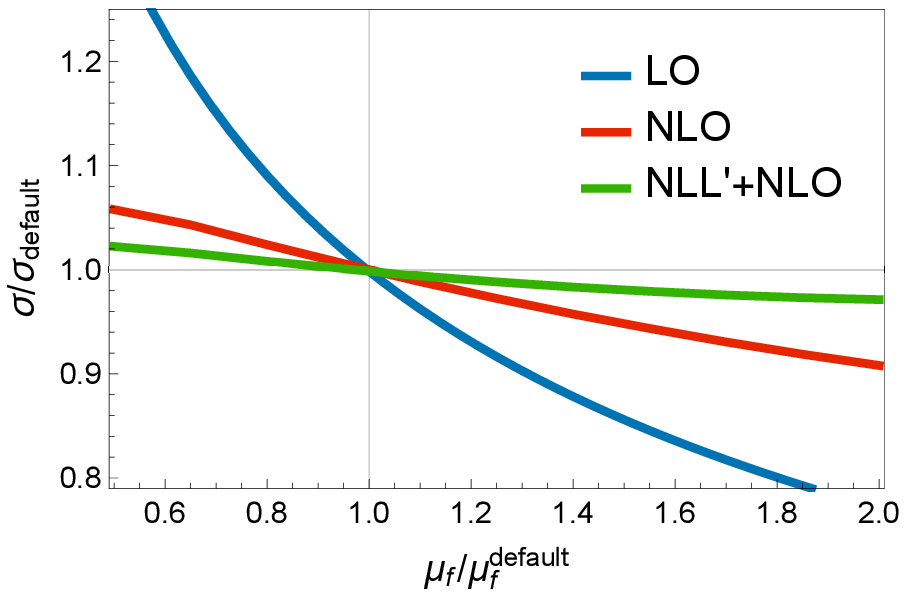}
\caption{\label{fig:scale_dep}The scale dependence of the total cross section at $\unit{13}{\TeV}$ LHC. The left plot shows the dependence of the NLL$'$+NLO result on the scales $\mu_h$, $\mu_s$ and $\mu_J$ entering the resummation formula. The right plot shows the dependence of the LO, NLO and NLL$'$+NLO results on the factorization scale $\mu_f$.}
\end{figure} 

We begin with the scale dependence of the total cross section at $\unit{13}{\TeV}$ LHC. The result at $\unit{14}{\TeV}$ LHC is similar and we do not show it here. The LO and NLO cross sections depend on the factorization scale $\mu_f$, where the strong coupling $\alpha_s$ and the PDFs are evaluated. The matched NLL$'$+NLO cross section depends in addition the hard scale $\mu_h$, the soft scale $\mu_s$ and the potential scale $\mu_J$. In the left plot of Fig.~\ref{fig:scale_dep}, we show the dependence of the NLL$'$+NLO cross section on $\mu_h$, $\mu_s$ and $\mu_J$. We observe that the dependence is rather mild. This can actually be expected since these scales only affects the region $\beta < \beta_{\text{cut}}$, which does not make dominant contributions. In the right plot of Fig.~\ref{fig:scale_dep}, we show the dependence of the LO, NLO and NLL$'$+NLO cross sections on the factorization scale $\mu_f$. It can be seen that the $\mu_f$ dependence is significantly reduced when going to higher orders in perturbation theory. At NLL$'$+NLO, the residue $\mu_f$ dependence is merely about 2\%. To estimate the theoretical uncertainties of the NLL$'$+NLO predictions, we vary the 4 scales up and down by a factor of 2, and add the resulting variations of the cross sections in quadrature.  

\begin{table}[t!]
\centering
\begin{tabular}{|c|c|c|}
\hline  & $\unit{13}{\TeV}$ LHC (pb) & $\unit{14}{\TeV}$ LHC (pb)
\\
\hline
NLO  & $0.493^{+5.8\%}_{-9.2\%}$ & $0.597^{+6.1\%}_{-9.2\%}$
\\
\hline
NLL$'$+NLO & $0.521^{+1.9\%}_{-2.6\%}$ & $0.630^{+2.3\%}_{-2.6\%}$
\\
\hline
$K$-factor & $1.06$ & $1.06$
\\
\hline
\end{tabular}
\caption{Results for the total cross section at NLO and NLL$'$+NLO accuracies. The uncertainties reflect scale variations only.}
\label{tab:xs}
\end{table}

The predictions for the total cross sections are summarized in Table~\ref{tab:xs}.
The $K$-factor is defined as the ratio of the NLL$'$+NLO cross section to the NLO one.
Although we do not expect huge effects from the resummation due to the $\beta^4$ suppression of the threshold region, we still find that the higher order threshold corrections enhance the cross section by 6\%. This should be taken into account for the high precision HL-LHC run. We also observe a significantly reduction of the scale dependence from NLO to NLL$'$+NLO. 
This should be contrasted to the result of \cite{Kulesza:2015vda}, where the NLL resummation was performed for the $\ln\beta$ corrections, while the $1/\beta$ corrections were added at fixed-order (i.e., not resummed).
The difference between our result and the result in \cite{Kulesza:2015vda} resides in the fact that we have computed the NLO hard function exactly for each color channel, and we have resummed the $1/\beta$ terms to all orders into the potential function.
The reduction of the scale dependence shows that these additional efforts are important phenomenologically.

\section{Conclusion}
\label{sec:conclusion}

In this work, we have generalized the resummation framework for $t\bar{t}$ production to investigate the associated production of a Higgs boson with a pair of top quarks. A major difference between the two processes is that for $t\bar{t}h$ production, the $t\bar{t}$ pair is recoiled by the Higgs boson and has a residue momentum of the potential scaling; while for $t\bar{t}$ production, the residue momentum of the top quark pair is of the ultrasoft scaling. The presence of the recoil momentum leads to several complications in the derivation of the factorization formula. We have shown that the next-to-leading power interaction between the ultrasoft mode and the potential mode does not contribute to the total cross section when the momentum of the Higgs boson is integrated over. We have also argued that the contributions from the potential mode can be resummed into a potential function, which is related to that in $t\bar{t}$ production via a boost. The final outcome of these considerations is that the total cross section for $t\bar{t}h$ production admits a similar factorization formula up to NLP as that for $t\bar{t}$ production. This similarity relies on subtle cancellations of the ultrasoft-potential interactions in the integrated cross section and on the same form of pNRQCD Lagrangians up to NLP, which may not hold at higher powers in $\beta$.

An important ingredient entering the factorization formula is the hard function, which was not known in the literature beyond the LO. We have explicitly calculated the NLO corrections to the hard function, decomposed into singlet and octet color configurations.
We have validated our factorization formula by expanding it to the NLO in $\alpha_s$ and comparing with the explicit calculations in \cite{Beenakker:2002nc}.
Based the factorization formula, we have derived a resummation formula at NLL$'$ accuracy using RG equations.
By matching to the NLO result, we are able to provide numeric predictions for the total cross sections at the NLL$'$+NLO accuracy.
We find that the resummation effects enhance the cross sections at $\unit{13}{\TeV}$ and $\unit{14}{\TeV}$ LHC by about 6\%, and reduce the scale dependence significantly.

We emphasize that the resummation framework in this paper can as well be applied to $t\bar{t}h$ production at a future $e^+e^-$ collider, where $\beta$ is fixed by the collider energy instead of being integrated over. The impact of resummation is expected to be more important in that case if the collider energy is not too far beyond the production threshold.

\section*{Acknowledgements}

We would like to thank Jan Piclum and Christian Schwinn  for helpful discussions.  This work was supported in part by the China Postdoctoral Science Foundation under Grant No. 2017M610685  and the National Natural Science Foundation of China under Grant No. 11575004 and 11635001.

\appendix
   
\section{Ingredients in the renormalization group evolution}
\label{app:coeff}

In this appendix, we list the ingredients entering the resummation formula \eqref{eq:resummed}. We begin with the QCD $\beta$-function, whose perturbative expansion is defined as
\begin{equation}
\beta(\alpha_s) = -2\alpha_s \sum_{n=0} \left( \frac{\alpha_s}{4\pi} \right)^{n+1} \beta_{n} \, ,
\end{equation}
where the one-loop and two-loop coefficients are given by \cite{vanRitbergen:1997va}
\begin{align}
\beta_{0} &= \frac{11}{3} C_A - \frac{4}{3} n_f T_F \, , \nonumber
\\
\beta_{1} &= \frac{34}{3} C_A^2 - \frac{20}{3} C_A n_f T_F - 4 C_F n_f T_F \, .
\end{align}
The evolution factor in the resummation formula is defined as
\begin{multline}
U_{ij}^\alpha(\mu_f,\mu_h,\mu_s) = \exp \bigg[ 2S^{\text{cusp}}_{ij}(\mu_h,\mu_s) - 2A_{ij}^{\text{cusp}}(\mu_s,\mu_f) \ln\frac{\mu_h}{\mu_s}
\\
+ 2A_{ij}^{\text{cusp}}(\mu_h,\mu_f) \ln\frac{\mu_h}{2m_t+m_h}
+ A^{s,\alpha}_{ij}(\mu_s,\mu_f) - A^{H,\alpha}_{ij}(\mu_h,\mu_f)
\bigg] \, ,
\end{multline}
where the functions $S^{\text{cusp}}_{ij}$, $A^{s,\alpha}_{ij}$ and $A^{H,\alpha}_{ij}$ are given by
\begin{align}
S^{\text{cusp}}_{ij}(\mu_a,\mu_b) &= -\int^{\alpha_s(\mu_b)}_{\alpha_s(\mu_a)} d\alpha_s \frac{(C_i+C_j)\gamma_{\text{cusp}}(\alpha_s)}{\beta(\alpha_s)}\int^{\alpha_s}_{\alpha_s(\mu_a)} \frac{d\alpha'_s}{\beta(\alpha'_s)} \, , \nonumber
\\
A^{k,\alpha}_{ij}(\mu_a,\mu_b) &= -\int^{\alpha_s(\mu_b)}_{\alpha_s(\mu_a)} d\alpha_s \frac{\gamma^{k,\alpha}_{ij}(\alpha_s)}{\beta(\alpha_s)} \, .
\end{align}

\section{New structures at next-to-next-to-leading power}
\label{app:nnlp}

In this work, we have only been concerned with up to next-to-leading power contributions. We have seen that at this order, the total cross sections for $t\bar{t}h$ production and $t\bar{t}$ production have a similar form of factorization in the threshold limit. However, this similarity is in general not expected to hold at higher powers in $\beta$. In this appendix, we discuss some next-to-next-to-leading power (NNLP) contributions which may lead to new structures in the factorization formula.

Firstly, the vanishing of the contribution from the $\vec{x} \cdot \vec{E}_{\text{us}}$ term strongly relies on the fact that at $\mathcal{O}(\beta)$, the integrand for the cross section must be linear in $\vec{p}_h$. This will no longer be true at NNLP, where one can have corrections quadratic in $\vec{p}_h$, which do not vanish even after integrating over all phase space. This may lead to new terms in the factorization formula which are not present in $t\bar{t}$ production.

Secondly, at NNLP the third-order terms in the effective Lagrangian will come into play. For example, the NNLP pNRQCD Lagrangian contains
\begin{equation}
\mathcal{L}_{\text{pNRQCD}}^{(2)}(x) \supset \int d^3\vec{r} \, \frac{\alpha_s}{2m_t^2} \big[ \mathcal{V}_{\text{nr}}(x,\vec{r}) + \mathcal{V}_{\text{rc}}(x,\vec{r}) \big] \, ,
\end{equation}
where 
\begin{align}
\mathcal{V}_{\text{nr}}(x,\vec{r}) &= \delta^{(3)}(\vec{r}) \, \psi^{\dag} T^{a} \psi \, \chi^{\dag} T^{a} \chi
+ \frac{1}{r} \Big[ \psi^{\dag} T^{a} \big( \vec{\nabla}^2_r \psi \big) + \big( \vec{\nabla}^2_r \psi^{\dag} \big) T^{a} \psi \Big] \chi^{\dag}T^{a} \chi \nonumber
\\
&+ \frac{3}{4r^3} \big( \vec{r}_i \vec{\partial}_r^j \psi^{\dag} \big) T^{a} \Big\{ [\sigma_i,\sigma_j] \psi \chi^{\dag} - \psi \chi^{\dag} [\sigma_i,\sigma_j] \Big\} T^{a} \chi \nonumber
\\
&+ \frac{r^2 \delta_{ij} - 3 \vec{r}_i \vec{r}_j}{8r^5} \psi^{\dag} T^{a} [\sigma_k,\sigma_i] \psi \chi^{\dag} T^{a} [\sigma_k,\sigma_j] \chi \, , \nonumber
\\
\mathcal{V}_{\text{rc}}(x,\vec{r}) &= -\frac{1}{r} \Tr \bigg\{ \chi \psi^{\dag} T^{a} \big[ \vec{\partial}^i_x (\vec{\partial}^i_r \psi) \chi^{\dag} \big] T^{a} -\big[ \vec{\partial}^i_x \chi (\vec{\partial}^i_r \psi^{\dag}) \big] T^{a} \psi    \chi^{\dag} T^{a} \bigg\} \nonumber
\\
&+ \frac{3}{4r^3} \Tr \bigg\{ \big( \vec{r}_i \vec{\partial}^j_x \chi \psi^{\dag} \big) T^{a} [\sigma_i,\sigma_j] \psi \chi^{\dag} T^{a} - \big( \vec{r}_i \vec{\partial}^j_x \chi \psi^{\dag} \big) T^{a} \psi \chi^{\dag} T^{a} [\sigma_i,\sigma_j] \bigg\} \, ,
\end{align}
where $\psi \equiv \psi(x^0,\vec{x}+\vec{r})$ and $\chi \equiv \chi(x^0,\vec{x})$.
Note that the term $\mathcal{V}_{\text{nr}}$ will also appear in $t\bar{t}$ production at NNLP, but the term $\mathcal{V}_{\text{rc}}$ is one power higher in $t\bar{t}$ production, since it involves the 3-momentum of the $t\bar{t}$ pair. The different counting of the $\mathcal{V}_{\text{rc}}$ term may spoil the simple relation \eqref{eq:Jrelation} at NNLP.

Finally, note that the above discussions are just speculations, and we cannot draw a definite conclusion unless we perform a more thorough analysis, which is beyond the scope of the current work.

\end{document}